\documentclass{aa}  

\usepackage{graphicx}
\usepackage{txfonts}
\usepackage{natbib}
\usepackage{subcaption}
\begin{document}

   \title{Host galaxy and orientation differences between different AGN types}
      \titlerunning{Host galaxy and orientation differences between different AGN types}

   \author{Anamaria Gkini 
          \inst{1,2},
          Manolis Plionis\inst{3,4},
          Maria Chira\inst{2,4}
          \and
          Elias Koulouridis\inst{2}
          }
   \authorrunning{A. Gkini, M. Plionis, M. Chira \& E. Koulouridis}
   \institute{ Department of Astrophysics, Astronomy \& Mechanics, Faculty of Physics, National and Kapodistrian University of Athens, Panepistimiopolis Zografou, Athens 15784, Greece
          \and
Institute of Astronomy, Astrophysics, Space Applications and remote Sensing, National Observatory of Athens, GR-15236 Palaia Pendeli, Greece
\and
National Observatory of Athens, GR-18100 Thessio, Athens, Greece
\and
Sector of Astrophysics, Astronomy \& Mechanics, Department of Physics, Aristotle University of Thessaloniki, Thessaloniki 54124, Greece
}


 
  \abstract
   {}
   {The main purpose of this study is to investigate aspects regarding the validity of the active galactic nucleus (AGN) unification paradigm (UP). In particular, we focus on the AGN host galaxies, which according to the UP should show no systematic differences depending on the AGN classification. }
   {For the purpose of this study, we used (a) the spectroscopic Sloan Digital Sky Survey (SDSS) Data Release (DR) 14 catalogue, in order to select and classify AGNs using emission line diagnostics, up to a redshift of $z=0.2$, and (b) the Galaxy Zoo Project catalogue, which classifies SDSS galaxies in two broad Hubble types: spirals and ellipticals.}
   {We find that the fraction of type 1 Seyfert nuclei (Sy1) hosted in elliptical galaxies is significantly larger than the corresponding fraction of any other AGN type, while there is a gradient of increasing spiral-hosts from Sy1 to LINER, type 2 Seyferts (Sy2) and composite nuclei.
These findings cannot be interpreted within the simple unified model, but possibly by a co-evolution scheme for supermassive black holes (SMBH) and galactic bulges.

Furthermore, for the case of spiral host galaxies we find the Sy1 population to be strongly skewed towards face-on configurations, while the corresponding Sy2 population range in all host galaxy orientation configurations has a similar, but not identical, orientation distribution to star-forming (SF) galaxies. These results also cannot be interpreted by the standard unification paradigm,  but point towards a significant contribution of the galactic disc to the obscuration of the nuclear region. This is also consistent with the observed preference of Sy1 nuclei to be hosted by ellipticals, that is, the dusty disc of spiral hosts contributes to the obscuration of the broad-line region (BLR), and thus relatively more ellipticals are expected to appear hosting Sy1 nuclei.}
   {}


   \maketitle
%
  
\section{Introduction}

The unification model of active galactic nucleus (AGN) explains the wide variety of features discerned in different classes of AGN in terms of the anisotropic geometry of the black hole's immediate surroundings  \citep[e.g.][]{ Antonucci1993, Urrya1995, Netzer2015}). It is well known that AGNs have a central supermassive black hole (SMBH) that accretes surrounding matter, and a toroidal structure by dust molecular gas  that absorbs a fraction of the emitted radiation. Near the SMBH, the high-speed gas clouds produce the broad-line region (BLR), while far away from this the torus clouds that move at low speeds produce the narrow-line region (NLR). In the simplest form of the unification paradigm (UP), the observed differences between type 1 Seyfert (Sy1) and type 2 Seyfert (Sy2) objects are due to orientation effects with respect to the line of sight to the observer.

An important implication of this simple model is that, since the observed differences between AGN types are attributed solely to orientation effects, the host galaxies of AGN should be intrinsically the same  \citep[e.g.][]{Netzer2015,Hickox2018}. Thus, despite their success in explaining a range of AGN observed features, such as the absence of broad-line features in the spectra of Sy2 galaxies, it has become clear that the simplest models of unification are inconsistent with observations and cannot explain aspects such as the lack of Sy1s in isolated pairs of galaxies \citep{Gonzales2008} and
in compact groups \citep{Martinez2008,Bitsakis2010}, the lack of hidden BLRs in many Sy2 AGNs in polarised spectra \citep{Tran2001T, Tran2003}, and the absence of detected BLRs in low-accretion-rate  AGNs \citep[e.g.][]{Nicastro2003,ElitzurHo2009}. Specifically, according to studies investigating the differences between Sy1s and Sy2s \citep[e.g.][]{Keel1980,Kinney2000,Koulouridis2006,Rigby2006,Martinez2006,Lacy2007,Lagos2011,Almeida2011,Elitzur2012,Villarroel2012,Koulouridis2014, Villarroel2014,Netzer2015,Villarroel2017,Bornancini2018,Zou2019,Yang2019,Bornacini2020,Malizia2020} and the connection with their host galaxies at redshifts $0.03<z<0.2$  \citep[e.g.][]{Villarroel2012,Koulouridis2013,Villarroel2014}, it has been shown that type 1 and type 2 Seyfert galaxies have different optical, mid-infrared, X-ray, and morphological properties \citep[e.g.][]{Sorrentino2006A,Slavcheva2011,Villarroel2017,Chen2017,Bornancini2018}, and they also reside in statistically different environments \citep[e.g.][]{Koulouridis2013,Villarroel2014,Jiang2016,Bornancini2017,Bornancini2018,Bornacini2020}. In particular, \citet{Bornacini2020} showed that the ultraviolet (UV), optical, and mid-infrared (IR) colour distribution of the different
AGN classes differ significantly, while \cite{Villarroel2014} found that
the activity and colour between the
neighbours of Sy1s and Sy2s are significantly different, and that the spiral
fraction of the host galaxies depends on the environment of
Sy1 and Sy2 galaxies in different ways. Moreover, it is found that the neighbours of Sy2 AGN are more star-forming and bluer than Sy1 AGNs \citep[see also][]{Koulouridis2013} and also that Sy2 hosts are surrounded by a larger number of dwarf galaxies \citep{Villarroel2012}. Additionally, the morphology of Sy1 galaxies shows no indications of close interactions, which means either that they rarely merge \citep{Koulouridis2014}, or that they are extremely short-lived AGNs \citep{Villarroel2012}. \citet{Bornancini2018} find that Sy2s have more abundant neighbours at high redshifts ($0.3<z<1.1$) and that Sy1 hosts are preferably elliptical or compact galaxies, while Sy2 hosts present a broader Hubble-type distribution. 

Many studies  \citep[e.g.][]{Almeida2011,Elitzur2012,Villarroel2017,Suh2019} reveal that differences between the two types of AGN also lie in the star formation rate (SFR), in the structure of the torus and the physics
of the central engines. Most notably, \citet{Suh2019} and \citet{Villarroel2017} found that the two Seyfert populations have different luminosities with \citet{Villarroel2017} supporting that type 1 AGNs are ten times more luminous than type 2s. \citet{Suh2019} also concluded that Sy1 and Sy2 host galaxies seem to have different SFRs, while \citet{Villarroel2017} detected many more supernovae (SNe) in type 2 AGN hosts, which implies differences between stellar ages and
more recent star formation in Sy2 hosts. On the other hand, \citet{Bornancini2018} found no difference between the SFR distribution of tracer galaxies
in both AGN samples at larger scales ($r_{p}<500$ kpc). In addition, in the context of differences in the structural features of the torus, \citet{Almeida2011} found that the Sy2 torus is broader and has more
clouds than that of Sy1 and that the optical depth of the clouds in Sy1 torus is larger than in Sy2. They suggest that the type 1/type 2 classification depends on the torus intrinsic properties rather than in the torus inclination \cite[e.g.][]{Elitzur2012,Audibert2017}. Previous work \citep[e.g.][]{Mendoza2015} has also claimed that the Sy1 and Sy2 active nuclei have different torus clumpiness, while they also found a dependence on whether the host is in a merger or isolated.

Since, observations do not fully comply with the predictions of the simple '{\it A type 2 AGN is just a type 1 AGN viewed through a dust}' model, alternative or complementary factors affecting the observed AGN types should be sought in order to explain the AGN variety.  Indeed, \citet{Koulouridis2006} and \citet{Jiang2016}, when studying the environments of Sy1 and Sy2 galaxies at low redshifts, found that both AGN classes have similar clustering properties \citep[e.g.][]{Zou2019}, but at scales smaller than $100 kpc$ Sy2s have significantly more neighbours than Sy1s  \citep[e.g.][]{Bornancini2018}. These results contrast with those of \citet{Melnyk2018}, who obtained similar local- and large-scale environments for the different AGN types.  \citet{Jiang2016} also found significant differences in the infrared colour distributions of the host galaxies of the two AGN types. Besides this,  \citet{Powell2018}
reported that nearby type 2 AGNs reside in more massive halos than
type 1s, which is unlike the results of \citet{Jiang2016}, who found that their halo masses are similar.

Furthermore, some studies \citep[e.g.][]{Keel1980,Maiolino1995,McLeod1995,Malkan1998,Matt2000,Schmitt2001,Martinez2006,Rigby2006,Lacy2007,Lagos2011,Goulding2012,Netzer2015,Burtscher2016,Jiang2016,Buchner2017,Lanzuisi2017,Bornancini2018,Zou2019,Malizia2020,Masoura2021}  claim that the host galaxy might have a non-negligible contribution to the optical obscuration of nuclei. For example, \citet{Zhicheng2018} suggest that obscuration on both circum-nuclear ($\sim$ pc ) and galactic ($\sim$ kpc) scales are important in shaping and orienting the AGN narrow-line regions. Further evidence comes from  high-resolution observations, in nearby type 2 AGNs, which have revealed that the
nucleus is covered by a large-scale – a few hundred pc – dust filament or diffuse dust lane \citep{Prieto2014}. Moreover, \citet{Chen2015} found that the obscuration at X-ray and optical bands in type 2 quasars is connected to the far-IR-emitting dust clouds usually located on the scale of the host galaxy. Furthermore, \citet{Malizia2020} recently published an analysis using a hard X-ray-selected AGN sample showing that material located in the host galaxy on scales of hundreds of parsec, while not aligned with the absorbing torus, can be extended enough to hide the BLR of some Sy1s causing their misclassification as Sy2 objects and giving rise to the deficiency of around 24\%  of Sy1s in edge-on galaxies. 

The different average AGN luminosities and stellar ages of the host
galaxies of the Sy1 and Sy2 populations could be explained within the framework of an evolutionary scenario. According to \citet{Koulouridis2006} and \citet{Krongold2002}, in some cases the interaction between gas-rich galaxies ignites starburst activity, while large amounts of gas and dust obscure the central nuclear region at this stage. As the starburst dies off, the remaining molecular gas and dust forms a torus around the disc and eventually the AGN will attenuate the obscuring medium. Namely, this model proposes an AGN evolutionary sequence going from starburst to type 2, and finally to type 1,  Seyfert galaxies  \citep[e.g.][]{Springel2005,Hopkins2006,Koulouridis2013,ElitzurTrump2014,Villarroel2014,Mendoza2015,DiPompeo2017a,Yang2019,Spinoglio2019}.

Studies in polarised light to Sy2 galaxies have shown that in many low-luminosity AGN the dusty torus is absent, while the BLR is also not detected \citep[e.g.][]{Elitzur2006, Perlman2007,Trump2011,Koulouridis2014,Hernandez2016}. These results are consistent with those of \citet{Trump2011},  who support the notion that above a specific accretion rate  ($(L/L_{\rm Edd}\gtrsim0.01)$) AGNs can be observed as broad-line or as obscured narrow-line AGNs, while for $(L/L_{\rm Edd}\lesssim0.01)$ the BLR becomes non-detectable. Additionally, the obscuring torus tends to become weaker or disappears \citep{ElitzurTrump2014}.

The above findings could be incorporated within the evolutionary scheme. If the accretion-rate-dependent scenario is valid, one would expect that AGNs could lose their torus or/and their BLR at the end of the AGN duty cycle, as the accretion rate drops below a critical value \citep[e.g.][]{ ElitzurHo2009,ElitzurTrump2014,Koulouridis2014}. This implies that firstly, AGNs can appear as type 1, after the quenching of the star-forming activity by the AGN feedback and the disappearance of the torus \citep{Krongold2002} and secondly, these type 1s will evolve to true type 2 AGNs due to the elimination of the BLR, based, for example, on the wind-disc scenario of \citet{ElitzurHo2009}. 

In summary, the plethora of results of many relevant studies clearly indicate that the viewing angle alone cannot fully account for the different AGN types. It is within this ideology that the current study lies, investigating the orientation properties of spiral hosts, as well as the Hubble-type distribution of different types of AGNs.

After the presentation of the data used in Sect. 2, the main body of our analysis is organised as follows. In Sect. \ref{AGN_type_morphologies}, we study the morphology frequency distribution of host galaxies for different AGN types comparing to that of the non-active star-forming sample, after statistically matching their respective redshift distribution in order to suppress possible evolutionary effects. 
In Sect. \ref{b_over_a}, we study the frequency distribution of spiral host galaxy orientations ($b/a$) for the Sy1 and Sy2 sub-samples, comparing with that of non-active star-forming (SF) galaxies (which we use as a control sample). We only used galaxies with high 'spirality' probability $>0.8$ (as defined by the Galaxy Zoo project), also statistically matching their respective redshifts and stellar mass distributions.

\section{Observational data}\label{data}

For the purposes of the current study, we used galaxy catalogues extracted from the Sloan Digital Sky Survey (SDSS) Data Release (DR) 14 \citep{Abolfathi2018} in five bands (u,g,r,i,z) with a  magnitude limit for the spectroscopic sample of $m_r = 17.77$ in the r-band, in order to have a homogeneous magnitude cut-off over the largest possible SDSS area. Our catalogues consist of $3.800$ Sy1, $56.846$ Sy2, $120.025$ Composite, $107.034$ LINER, and $263.223$ SF galaxies with redshift $z<0.2$. The above classifications are derived from the spectroscopic reanalysis by the MPA-JHU teams of a large sub-sample of SDSS galaxies, as described in \citet{Tremonti2004} and \citet{Brinchmann2004}. The morphology characterisation of the galaxies is based on the Galaxy Zoo Project \citep{Lintott2008}, a crowd-sourced astronomy project that asks citizens to characterise galaxies as spirals or ellipticals and to determine the rotation direction of spirals by inspecting SDSS galaxy images \citep{Raddick2007}. In order to ensure the reliable classification of our galaxies, we performed a quality cut on our catalogue, rejecting galaxies with Ha rest-frame equivalent width (EW) $<8$ \AA, which, although it is a rather arbitrary limit, we verified to be adequately secure via an inspection of a sufficiently large sub-sample.

The Sy1 sample (see Appendix) comprises all galaxies with a Balmer line width of $\sigma$ greater than 500 km/sec (FWHM>1180 km/s). We note that in the SDSS database all such sources are catalogued with $\sigma$=500 km/sec, while this is actually a lower limit. Visual inspection of a large number of spectra validated that these sources are bona fide broad-line Seyferts. All spectra with an emission line of $\sigma>$200 km/sec are also characterised as broad line in the SDSS database. We visually reviewed all spectra in our spiral galaxy sample that fall in this category (450 sources), and we conclude that they contain very few that could be unambiguously classified as broad-line AGNs. Therefore, to reduce noise in our analysis, we chose to exclude all spectra with Balmer lines of $\sigma<$500 km/sec from
our Sy1 sample. 

After the above procedures, our final sample of secure objects consists of 1.378 Sy1, 7.498 Sy2, 26.544 Composite, 1.926 LINER, and 203.298 SF galaxies.

The classification between different type of narrow-line AGNs and SF galaxies has been performed utilising the BPT diagram classification method of \citet{Baldwin1981}.
 
It is important to note that we will use the SF galaxies, of which the vast majority are spirals, as our control sample since their disc orientations cover the whole range of viewing angles with respect to the line of sight. Additionally, they are mainly non-AGN galaxies, and even when hosting an AGN in their centres the star formation dominates the emission. This leads to spectra characteristic of non-AGN galaxies \citep{Siebenmorgen2015}. 

Before proceeding, we felt it necessary to investigate the level of consistency between the morphology classification of the Galaxy Zoo project and our own assessment via an inspection of the host galaxy image.
To this end, we selected a small sub-sample of images and spectra of galaxies of the highest 'spirality' (probability of being a spiral) and 'ellipticity' (probability of being an elliptical). In detail, we selected 50 nearby galaxies, five spirals, and five ellipticals of every AGN type, with redshift $z<0.1$ in order to ensure a high-quality image. We concluded that although the Galaxy Zoo morphology classification is reliable for the majority of our galaxies, we found some cases in which the assessment of the citizens participating in the Galaxy Zoo project seems to be dubious. Our analysis showed that we can trust the characterisation of spirals up to $z\sim 0.2$ while that of ellipticals only up to $z\sim 0.1$ since the spiral features are more difficult to distinguish at higher redshifts, where a galaxy image with weak spirality can be interpreted as being an elliptical.

\section{Methodology and results}
 This section is organised as follows. In Sect. \ref{AGN_type_morphologies} we study the Hubble-type frequency distributions of the host galaxies for the different samples: Sy1, Sy2, LINER, Composite, or SF, while in Sect.  \ref{b_over_a} we compare the projected axial-ratio (related to the orientation with respect to the line of sight) frequency distributions for the spiral hosts of Sy1s and Sy2s.

\subsection{Hubble-type distribution}\label{AGN_type_morphologies}
In this section, we seek to reveal if there is any correlation between the AGN type and the Hubble-type morphology of their host galaxies. According to the simplest unification scheme, the different AGN classes are a result of different viewing angles with respect to the orientation of an obscuring torus, and thus the properties of the host galaxies should not show any statistically significant differences. For this study, we used sub-samples of Sy1, Sy2, LINER, Composite, and SF galaxies, derived from our SDSS catalogue with a redshift limit of $z<0.1$, in order to obtain more robust Hubble-type classification (as discussed in Sect. \ref{data}). For these sub-samples, we generated the frequency distribution of the Zoo 'ellipticity' and 'spirality' probabilities, which within the context of the  simplest unification scheme, are expected to be statistically the same for all AGN classes \citep[e.g.][]{Antonucci1993,Urrya1995}.

\begin{figure}
\includegraphics[width=0.5\textwidth]{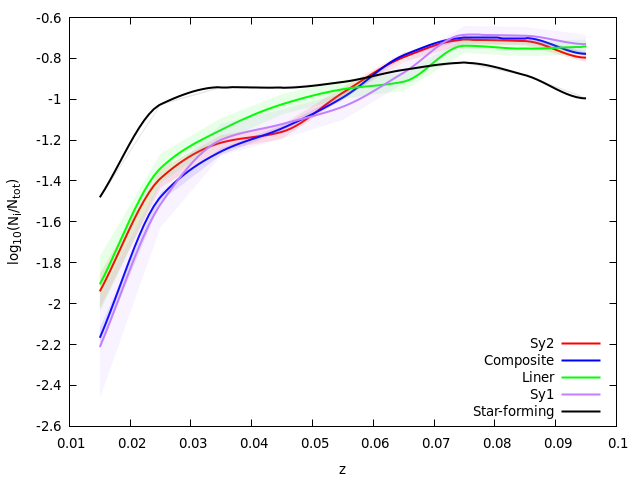}    
\includegraphics[width=0.5\textwidth]{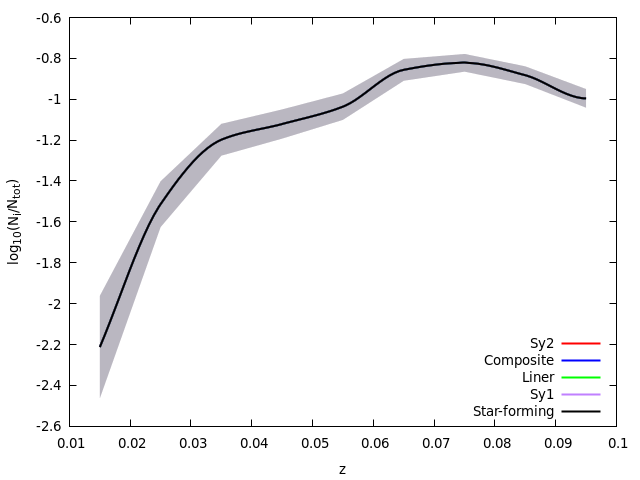}    
\caption{{\em Upper panel:} Normalised redshift frequency distributions for the four sub-samples of AGN types and SF galaxies, limited to $z<0.1$. The different sub-samples are colour-coded  as denoted in the key. {\em Lower panel:} Redshift-matched normalised distribution. The shaded area corresponds to the 1$\sigma$ Poisson uncertainty.}
\label{RedGa}
\end{figure}

In order to be able to compare the different probability distributions, avoiding possible evolutionary effects, it is necessary to take into account any statistically significant differences between the redshift distributions of the different samples, which we present in Fig.\ref{RedGa}-({\em upper panel}). Due to the different number of objects in the sub-samples, and in order to reveal systematic trends among the different sub-samples, we normalised the distributions dividing with the total number of objects in each sub-sample. Using a random sampling procedure, we matched the normalised redshift distributions to a common fractional distribution (Fig.\ref{RedGa} - {\em lower panel}). Because of the low number of Sy1 galaxies and in order to avoid further depleting it, we re-sampled all other activity types, so that their normalised distributions are matched to that of the Sy1 sample.  

\begin{figure}
\includegraphics[width=0.5\textwidth]{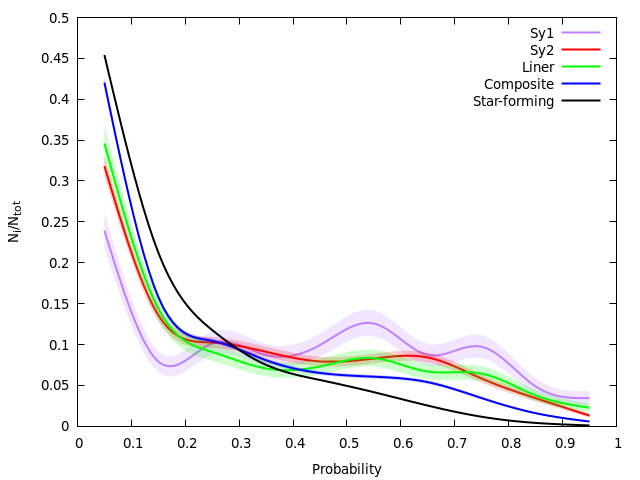}    
\includegraphics[width=0.5\textwidth]{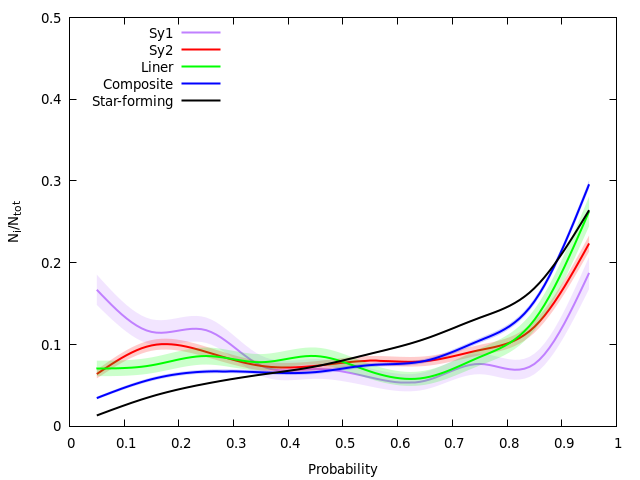}   
\caption{Probability distributions of ellipticity ({\em upper panel}) and spirality ({\em lower panel}) for the Sy1, Sy2, LINER, Composite and SF host galaxies, limited to $z<0.1$. The different AGN and SF sub-samples are colour-coded as in Fig.\ref{RedGa}, while the shaded area corresponds to the 1$\sigma$ Poisson uncertainty.}
\label{ESprob}
\end{figure}

For the redshift-matched sub-samples, in Fig. \ref{ESprob} we present the ellipticity ({\em upper panel}) and spirality ({\em lower panel}) probability distributions. The two quantities are complementary, which is to be expected since the sum of the two Zoo probabilities should be roughly equal to unity. Moreover, we should note that SF galaxies, with spectra dominated by young stellar populations, are by definition spirals \citep{Hubble1926}, and indeed, as seen in Fig. \ref{ESprob}, their spirality distribution (black solid line) peaks at high-spirality probabilities, while their respective ellipticity distributions, which are roughly complementary, peak at  zero-ellipticity probabilities. The SF galaxy sample can thus be used as a control sample of the Zoo Project spirality and ellipticity distributions of the various AGN host galaxies. 

In Fig. \ref{ESprob}, we see that the different AGN classes are distributed in the whole range of probabilities indicating a wide range of Hubble-type hosts, with the predominance of spirals. However, the normalised frequency distribution of the different AGN types are dissimilar at a statistically significant level (as indicated by the 1$\sigma$ Poisson uncertainty), which implies a different Hubble-type distribution for the different AGN types, a result that contradicts the original unification paradigm according to which there should be no dependence of the AGN class on the host galaxy's Hubble-type classification \citep[e.g.][]{Antonucci1993}.
Interestingly, the Sy1s show a peak at both high- and low- spirality probabilities, indicating a relatively higher fraction of Sy1s, with respect to other AGN types, residing in elliptical hosts.

\begin{figure}
\includegraphics[width=0.5\textwidth]{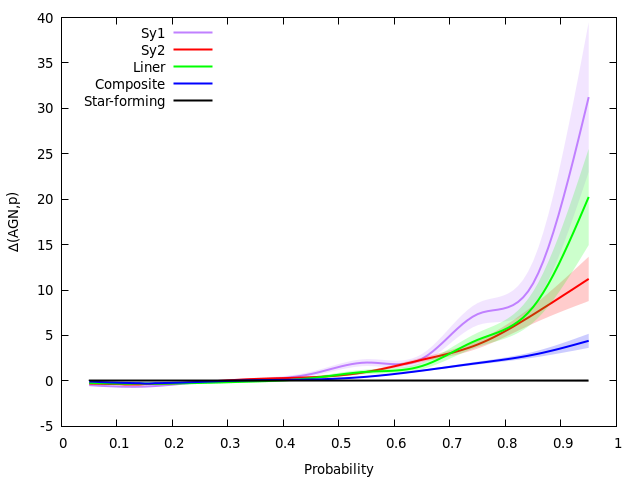}
\includegraphics[width=0.5\textwidth]{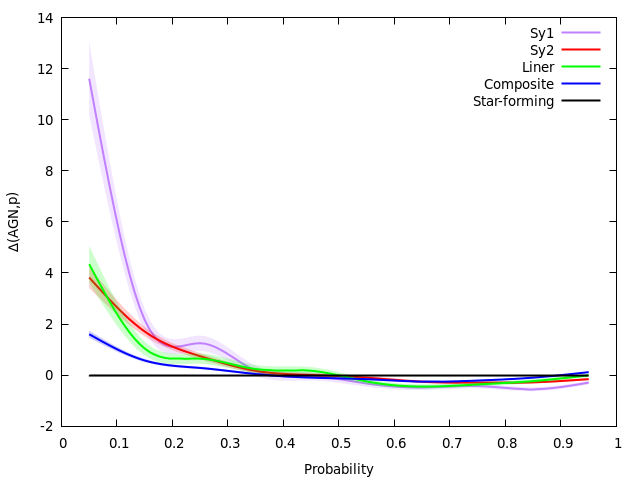}    
\caption{Values of the factor $\Delta({\rm AGN, p})$ by which the different AGN types exceed the corresponding fractional number of SF galaxies for the ellipticity ({\em upper panel}) and spirality (({\em lower panel}) cases. The different AGN and SF sub-samples are colour-coded as in Fig.\ref{RedGa}, while the shaded area corresponds to the 1$\sigma$ Poisson uncertainty.}
\label{NESprob}
\end{figure}

For a more revealing comparison of the previously discussed morphology difference of the various AGN host galaxies with respect to SF galaxies, in Fig.\ref{NESprob} we present the excess factor by which the fractional number of the various AGN types exceeds that corresponding to SF galaxies, for each spirality or ellipticity probability:
$$\Delta(\rm AGN, p)=\frac{\rm N_{i}(AGN)/N_{tot}(AGN)}{\rm N_{i}(SF)/N_{tot}(SF)} - 1\;.$$

Upon inspecting Fig.\ref{NESprob}, it becomes evident that Sy1s show the highest relative preference for elliptical hosts with respect to LINERs, Sy2s, and Composites, although all AGN types, each at a different degree, appear in elliptical hosts.

\subsection{Effect of the galactic disc on the obscuration of the AGN}\label{b_over_a}
We wish to test the hypothesis that the AGN host galaxy contributes to the obscuration of the AGN emission. To this end, we studied the orientation of spiral galaxies hosting Sy1 and Sy2 nuclei, limited to $z<0.2$. We only selected spiral hosts for this test since the orientation of spirals with respect to the line of sight can be quantified via their projected axis ratio, $(b/a)$. Furthermore, in order to reduce noise, we required a high Galaxy Zoo spirality probability, that is $p>0.8$. We also used the corresponding sub-sample of SF galaxies as a reference (or control) sample.  

\subsubsection{Matching the redshift distributions}\label{Redshift}
\begin{figure}
\includegraphics[width=0.5\textwidth]{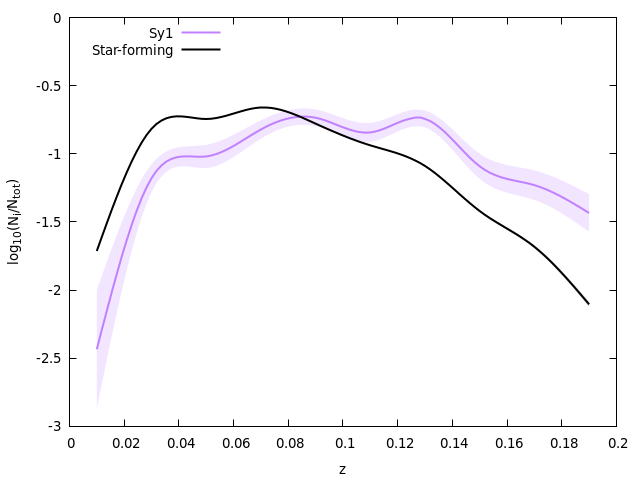}\hfill
\includegraphics[width=0.5\textwidth]{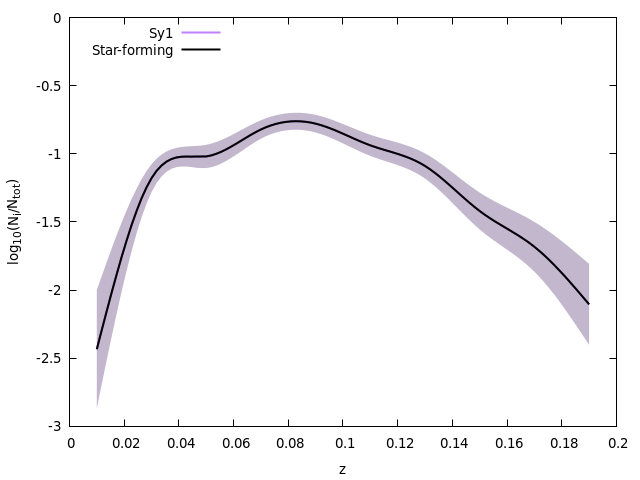}
\caption{{\em Upper panel:} Normalised redshift distribution of spiral Sy1 and SF galaxies.  {\em Lower panel:}  Normalised redshift-matched distribution of Sy1s and the corresponding SF control sample. The Sy1 and the SF sub-samples are colour-coded as in Fig.\ref{RedGa}, while the shaded area corresponds to the 1$\sigma$ Poisson uncertainty.}
\label{RBS}
\end{figure}

\begin{figure}
\includegraphics[width=0.5\textwidth]{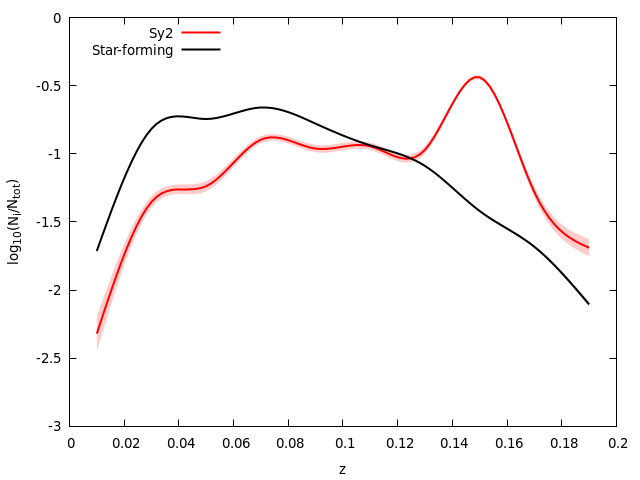}
\includegraphics[width=0.5\textwidth]{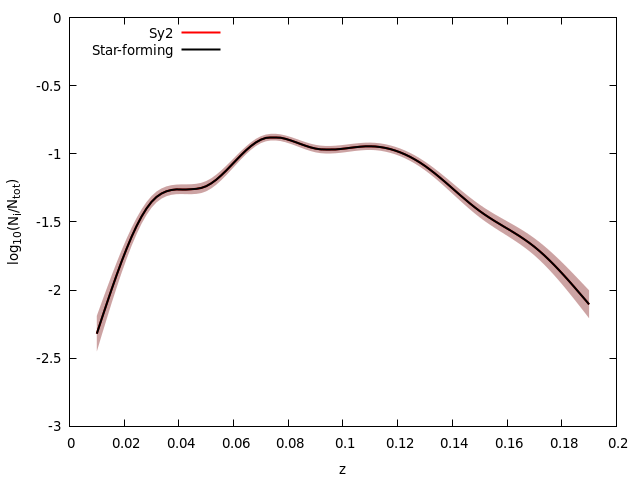}
\caption{{\em Upper panel:} Normalised redshift distribution of spiral Sy2 and SF galaxies. {\em Lower panel:}  Normalised redshift-matched distribution of Sy2 and the corresponding SF control sample. The Sy2 and the SF sub-samples are colour-coded as in Fig.\ref{RedGa}, while the shaded area corresponds to the 1$\sigma$ Poisson uncertainty.}
\label{RSS}
\end{figure}

In Figs.\ref{RBS}-\ref{RSS}-({\em upper panel}) ,we present the Sy1 and Sy2 sub-sample normalised redshift distributions and compare them with the corresponding distribution of SF galaxies. As can be clearly seen, and also as quantified by the Kolmogorov-Smirnov (KS) test (p-value $\sim$ $0$ and  $\sim$ $10^{-4}$, respectively), the distributions are significantly different. For $z\lesssim 0.08$, the fraction of both the AGN populations is lower than that of the SF galaxies, while for higher values, $z\gtrsim 0.1$, the AGN fractions are greater than that of SF galaxies. 
Whether due to observational biases or evolutionary effects, understanding such differences is out of the scope of the current work. However, we needed to ensure that our results would not be affected by such biases and thus followed the same re-sampling technique as in Sect.  \ref{AGN_type_morphologies}, in order to obtain matched-redshift distributions, which are shown in Figs. \ref{RBS}-\ref{RSS} -({\em lower panel}).

We can now proceed to a meaningful comparison of the orientation distribution of Sy1s and Sy2s with respect to the SF case. We expect that, according to the simplest unification model, the orientation distributions of Sy1s and Sy2s hosted in spiral galaxies must be identical to that of spiral SF galaxies. In presenting our results, we again normalise the distributions by the total number of objects in each sub-sample. 

\begin{figure}
\begin{center}
\includegraphics[width=0.5\textwidth]{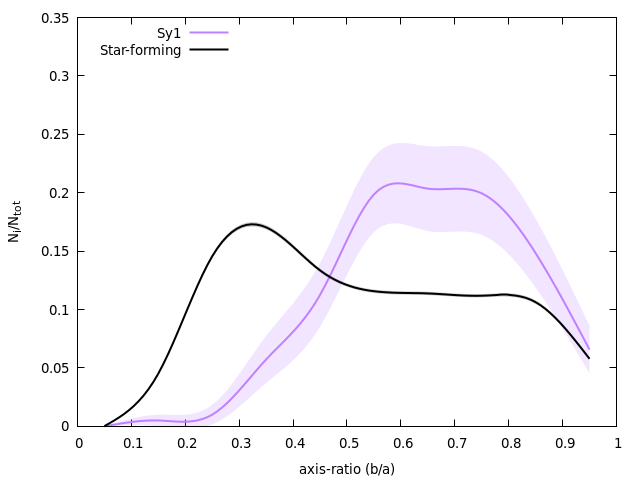}
\end{center}
\caption{Distribution of the projected axial ratio for spiral galaxies hosting Sy1s and the corresponding control sample of SF galaxies. The Sy1 and the SF sub-samples are colour-coded as in Fig.\ref{RedGa}, while the shaded area corresponds to the 1$\sigma$ Poisson uncertainty. }
\label{OBS}
\end{figure} 

In Fig.\ref{OBS}, we present the comparative plot of the normalised axis-ratio ($b/a$) distributions for the spiral galaxies hosting Sy1 nuclei (purple) and for the control sample of SF galaxies. It is evident that the two distributions are different, which we also confirm with a KS test (p-value $\sim10^{-8}$); the $b/a$ distribution of Sy1 is skewed towards high $b/a$ values, indicating inclination angles closer to face-on orientations, while the control sample covers the full range of orientation angles. The distribution of the control sample shows a peak at $b/a\approx 0.35$, whereas the Sy1 population peaks at $b/a\approx 0.7$. Moreover, for $b/a<0.4$ we find that the fraction of Sy1 host galaxies decreases dramatically compared to SF galaxies. Thus, we conclude that the type 1 Seyferts tend to be more 'face-on' compared to SF galaxies, and only a very small fraction of Sy1 galaxies are found to have $b/a$ values close to the edge-on orientation.

\begin{figure}
\begin{center}
\includegraphics[width=0.5\textwidth]{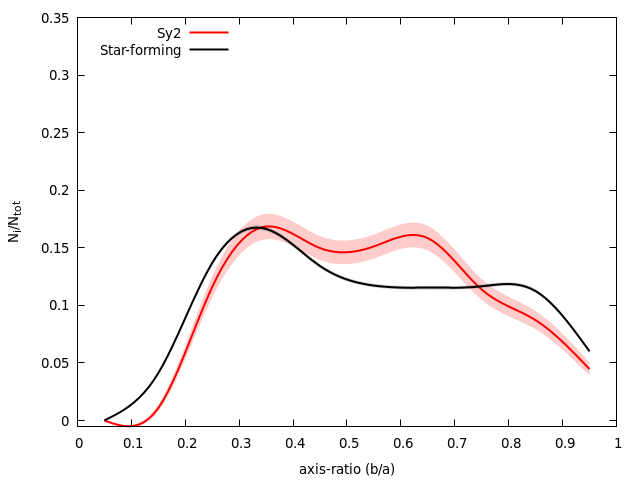}
\end{center}
\caption{Distribution of projected axial ratios for spiral galaxies hosting type 2 AGNs and the corresponding control SF sample. The Sy2 and the SF sub-samples are colour-coded as in Fig.\ref{RedGa}, while the shaded area corresponds to the 1$\sigma$ Poisson uncertainty. }
\label{OSS}
\end{figure}

In Fig. \ref{OSS}, we present the respective $b/a$ comparison plot for the Sy2 and the star-forming subsamples, and we find that although their distributions are significantly more similar than the corresponding of Fig.\ref{OBS}, they are still statistically different as confirmed by a KS test (p-value $\sim10^{-11}$). In detail, both the type 2 Seyferts and the SF galaxies are distributed in the whole range of $b/a$, with the star-forming peaking at b/a$\approx 0.35$, while the Sy2 appear to have two local maxima, at b/a$\approx 0.37$ and at b/a$\approx 0.65$. A significantly higher fraction of Sy2 galaxies has values of $b/a\in(0.3,0.8)$, while for values $b/a < 0.3$ (close to edge-on orientations) and $b/a > 0.8$ (close to face-on orientations), the fraction of Sy2s is lower than that of SF galaxies.

\subsubsection{Matching stellar mass distributions}\label{stellar}

 Comparing AGN samples of different intrinsic luminosity distributions may introduce a bias to the results. To this end, one could use physical properties that are proxies of the AGN intrinsic luminosity, as the luminosity of [O III]5007 emission line or the stellar mass of the host galaxies. However, since we are studying the additional absorption by the host galaxy that may also obscure the NLR, the use of the [OIII] line can bias our results. Therefore, we chose to match the stellar mass distributions of our samples, derived by the MPA-JHU teams, using the methodology defined in \citet{Kauffmann2003} and applied to photometric data as described in \citet{Salim2007}. Furthermore, this normalisation eliminates any possible bias introduced by the excessive absorption in one of the samples because of the uneven distribution of massive hosts.  

\begin{figure}
\includegraphics[width=0.5\textwidth]{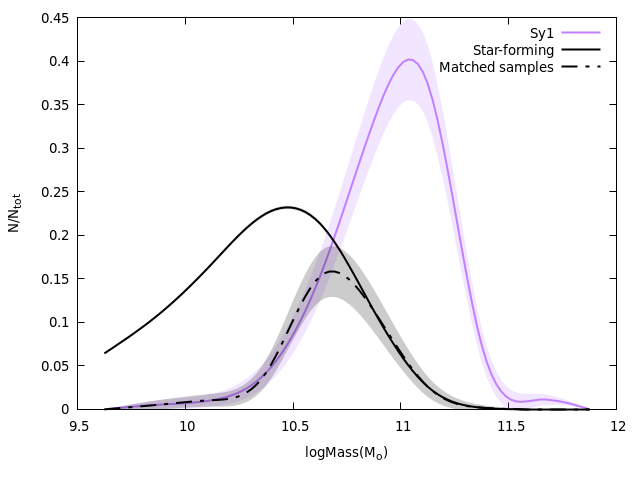}
\includegraphics[width=0.5\textwidth]{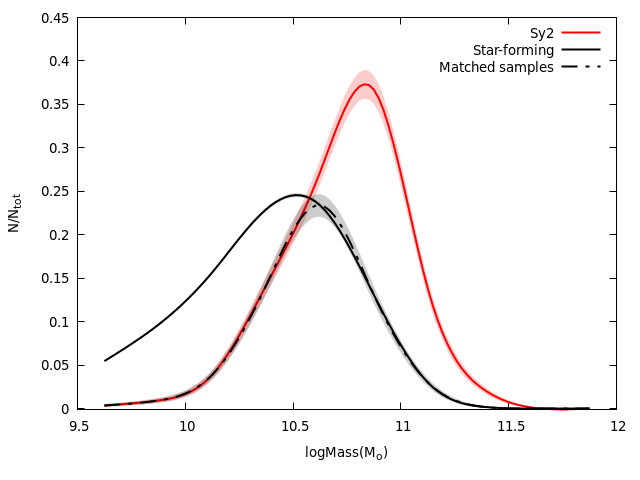}
\caption{Normalised stellar-mass distribution of spiral Sy1 ({\em upper panel}) and Sy2 ({\em lower panel}) with the corresponding SF and stellar-mass-matched sample (dashed line). The Sy1, Sy2, and SF sub-samples are colour-coded as in Fig.\ref{RedGa}, while the shaded area corresponds to the 1$\sigma$ Poisson uncertainty.}
\label{SM}
\end{figure}

For the already redshift-matched samples, in Fig.\ref{SM} we present the normalised stellar mass distributions of type 1 ({\em upper panel}) and type 2 ({\em lower panel}) AGNs with the corresponding distribution of SF galaxies. It is evident, but also confirmed by a KS test, that the distributions are statistically different (p-value $\sim 0$). We find that both AGN classes are identified with more massive host galaxies compared to the sample of SF galaxies, a result which is in agreement with that of \citet{Bornacini2020}.
Even though the direct comparison of the Sy1 and Sy2 stellar mass distributions is out of the scope of this work, it is noteworthy to mention that many previous studies \citep[e.g.][]{Yang2017,Yang2019} have shown that both AGN types have similar stellar mass hosts (see discussion in \citet{Zou2019}), which we also confirmed by a simple inspection of the corresponding figures.

In order to acquire more robust results and to ensure the removal of possible luminosity biases we follow the same re-sampling technique, as in Sect. \ref{AGN_type_morphologies}, to obtain matched stellar-mass distributions, which are shown in Fig. \ref{SM} (dashed distributions), at the expense, however, of significantly reducing the size of the AGN samples. This also excludes the most massive host galaxies from our samples, which could indeed reduce the significance of the effect (galaxy-disc obscuration of nuclear region) that we are seeking to investigate.

\begin{figure}
\begin{center}
\includegraphics[width=0.5\textwidth]{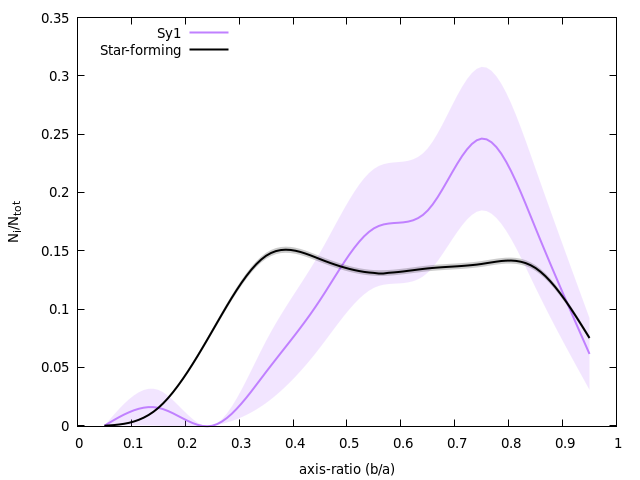}
\end{center}
\caption{Distribution of the projected axial ratio for stellar-mass-matched spiral galaxies hosting Sy1s and the corresponding control sample of SF galaxies. The Sy1 and SF sub-samples are colour-coded as in Fig.\ref{RedGa}, while the shaded area corresponds to the 1$\sigma$ Poisson uncertainty. }
\label{OSMBS}
\end{figure}

We can now detail the axial-ratio (b/a) distributions for the spiral-redshift- and stellar-mass-matched sub-samples of Sy1, Sy2, and star-forming galaxies. In Fig.\ref{OSMBS}, we present the normalised b/a comparative plot for Sy1 galaxies and the corresponding control sample. The KS test confirmed that the two distributions are different (p-value $\sim 10^{-3}$). Compared to Fig.\ref{OBS}, despite the reduction of the Sy1 sample size, both plots show similar behaviours, confirming that the results obtained in Sect. \ref{Redshift}  are robust.

\begin{figure}
\begin{center}
\includegraphics[width=0.5\textwidth]{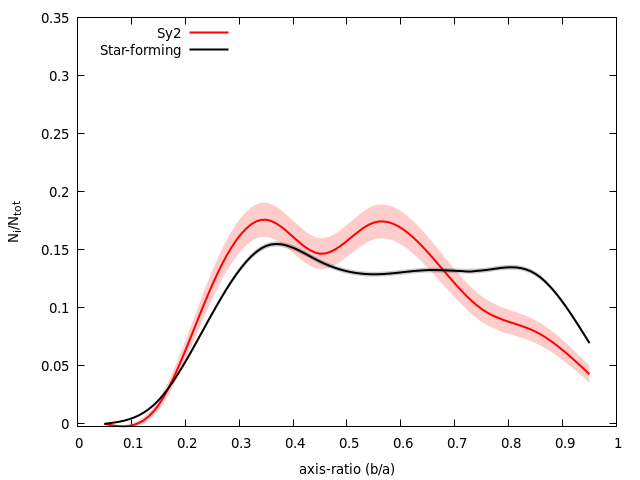}
\end{center}
\caption{Distribution of the projected axial ratio for stellar-mass-matched spiral galaxies hosting Sy2s and the corresponding control sample of SF galaxies. The Sy2 and SF sub-samples are colour-coded as in Fig.\ref{RedGa}, while the shaded area corresponds to the 1$\sigma$ Poisson uncertainty. }
\label{OSMSS}
\end{figure}

Correspondingly, in Fig.\ref{OSMSS} we present the normalised comparative plot for Sy2 and SF galaxies, and again the resulting distributions are similar to those of Fig.\ref{OSS}, again confirming that the main results presented in Sect. \ref{Redshift} are robust. However, in this case there is a small but distinct difference at small b/a values with the fraction of Sy2 being lower than that of SF for $b/a<0.2$, in contrast with Fig.\ref{OSS} where this transition occurs for $b/a\approx0.3$. Thus, when matching to stellar mass (Fig.\ref{OSMSS}), the fraction of Sy2 with small b/a is not significantly different to that of SF galaxies (Fig.\ref{OSS}).

\section{Discussion}
\subsection{Why do Sy1 nuclei favour elliptical hosts?}

There are various indications for differences in the host galaxies of AGNs in the literature. Our results in the low-redshift regime are in agreement with the results of \citet{Bornancini2018}, based on higher redshift $(0.3<z<1.1)$  type 1 and type 2 AGNs  from the COSMOS2015 catalogue, who showed that the type 1 AGN host galaxies appear more elliptical and compact than those of type 2 AGNs that span the whole spiral to elliptical Hubble-type range. In addition, using data from the fourth SDSS Data Release (DR4), \citet{Sorrentino2006A}  found that 76\% of type 1 Seyfert host galaxies are elliptical, while the corresponding ratio of Seyfert 2 hosted in early-type galaxies is 56.8\%. \citet{Slavcheva2011} found that more type 1 than type 2 AGNs prefer elliptical hosts. Similarly, \citet{Villarroel2017} and \citet{Chen2017} concluded that Sy2 nuclei reside more in spiral hosts ($\sim30-40\%$) than Sy1 nuclei do ($\sim20\%$), in disagreement with the simplest unification model.

The different bulge distributions of Sy1 and Sy2 might be related to an evolutionary sequence of AGN activity  \citep[e.g.][]{Krongold2002,Tran2003,Koulouridis2006,Villarroel2012,Koulouridis2013,Villarroel2014}. These results indicate a possible co-evolution scheme between galaxies and SMBHs  \citep[e.g.][]{Springel2005,Hopkins2006,Hopkins2008a,Hopkins2008b,Madau2014} \citep[see also reviews by][]{Kormendy2013,Heckman2014}. 

At the initial merging phase, during the enhanced star-forming activity and accretion, the AGN is mostly obscured because of the large amount of gas and dust in the circum-nuclear region. However, the AGN will eventually be revealed after the surrounding material is consumed, or expelled by radiation pressure. In parallel to the AGN evolution, the stellar population of the merger will also rapidly redden and the host will eventually be transformed to a quiescent elliptical galaxy \citep[e.g.][]{Hopkins2008a,Hopkins2008b}. AGN feedback plays an essential role in both AGN and galaxy transformation and probably leads to the observed SMBH-bulge relation \cite[e.g.][]{Magorrian1998,Ferrarese2000, Gebhardt2000}. Although, major mergers and powerful AGNs are far less common in the local Universe, compared to redshift $z>1$, the difference found in the current work could be partly due to the co-evolution of the bulge and AGN activity after such events.  

Alternatively, or in addition, the difference could also be due to the obscuration of the BLR from the gas- and dust-rich disc of a spiral galaxy, especially in edge-on systems. This possibility is discussed further in the next section.

\subsection{Why do Sy1 nuclei favour face-on orientations of the host galaxy?}

Our analysis also showed that the orientations of the Sy1 and Sy2 spiral host galaxies are significantly different when compared to the control sample of the spiral SF galaxies. This is also unexpected within the  simplest unification scheme, and the interpretation of our results points towards two possible scenarios: 
\begin{enumerate}
    \item additional obscuration of the AGN by dust in the galactic disc near the circum-nuclear region
    \item some level of statistical co-alignment of the plane of the torus and that of the galactic disc. 
\end{enumerate}

According to the latter scenario, Sy1 galaxies, which have a 'face-on' torus, also have more frequently face-on host galaxy orientations. However, if that is the case, we should correspondingly expect a higher fraction of 'edge-on' host galaxies in the Sy2 distribution with respect to SF galaxies, but this is not observed. Without excluding this scenario, since processes like merging may induce a rearrangement and facilitate such a co-planarity, we do not have a clear indication to confirm this hypothesis (see, however, \citet{Maiolino1995} for a relevant discussion).

Our results favour the scenario where additional obscuration, caused by the host galaxy, might affect the classification of Seyfert types, regardless of the inclination of the torus, as also claimed by \citet{Lagos2011}.  This result is similar to the one found by \citet{Keel1980}, who was the first to discover a deficiency of Sy1 galaxies in edge-on hosts \citep[see also][]{Lawrence1982}. Recent works support the idea that galactic-scale dust is responsible for additional obscuration \citep[e.g.][]{Burtscher2016}.  \citet{Buchner2017} showed that a large fraction (up to $40\%$) of the obscuration observed in AGNs, at least in the Compton-thin
regime, is not due to the nuclear torus, but to the galaxy-scale gas in the host, while \citet{DiPompeo2017b} estimated the fraction of IR-selected type 2 AGNs that are obscured by dust outside the torus to be $\sim 25\%$. Also, using data from the intermediate Palomar Transient Factory, the SDSS DR7, and Galaxy Zoo, \citet{Villarroel2017} found that for the majority of Sy2 galaxies (up to $90\%$), the obscuration must stem from larger scales: host galaxy obscuration and/or a large-scale environment in which the host galaxies reside.

Furthermore, having studied a sample of nearby Compton-thick AGNs, \citet{Goulding2012} concluded that the dust of the host galaxy, and not necessarily the compact torus, is the dominant obscurer of the central engine. Using Spitzer data of six $0.3<z<0.8$ type 2 quazars, \citet{Lacy2007} found a contribution of the extinction towards the nucleus from an extended star-forming disc on scales of kiloparsecs, in addition to, or instead of, the traditional dusty torus. Moreover, also using high-redshift type 2 quasars from Spitzer and VLA data of the Spitzer First Look Survey, \citet{Mart2006} concluded that the nuclear region could be effectively obscured by dust on large scales, away from the torus. Additionaly,  using X-ray-selected AGNs with spectroscopic redshifts in the Chandra Deep Field South (CDFS), \citet{Rigby2006} argued that part of the column density that obscures the soft X-rays may come from the galactic disc. Last but not least, \citet{Malizia2020} used the hard X-ray-selected sample of AGNs detected by INTEGRAL/IBIS to show that material located in the host galaxy on scales of hundreds of parsecs and not aligned with the absorbing torus can sufficiently hide the BLR of some type 1 AGNs, causing their classification as type 2 objects and giving rise to the deficiency of type 1 in edge-on galaxies.

Although the deficiency of edge-on Sy1 galaxies is unexpected in the simplest unification model, it does not necessarily contradict it. It might be assumed that in the case of edge-on
Seyfert galaxies the gas and dust along the host galaxy disc can act in the same way as the torus, blocking
the direct view of the BLR, thus leading to a
classification as a Sy2 galaxy \citep[e.g.][]{Schmitt2001}.

\subsection{Why there is a deficit of Sy2 nuclei in edge-on and face-on spiral hosts with respect to SF galaxies?}

Regarding the small but significant difference in the $b/a$ distributions of the spiral Sy2 and SF galaxies (Fig.\ref{OSS}), at low $b/a$ values a possible explanation could be that in extreme edge-on orientations the dust of the galactic disc can obscure not only the BLR but also the NLR region. In an early study, \citet{McLeod1995} used samples of optically and soft-X-ray-selected Seyferts and found a bias against inclined spiral hosts, while hard-X-ray-selected samples were found unbiased. In addition, \citet{Malkan1998} argued that in Sy2 galaxies irregular structures at large distances can provide sufficient absorbing column density for the nuclear source. Similar results were presented in \citet{Rigby2006}, where they attributed the appearance of X-ray-selected AGNs as optically 'dull' to galactic absorption.

On the other hand, there is a hypothesis that low luminosity Seyferts may be diluted by high-luminosity host galaxies, in which the continuum can hide the AGN lines. This may be the case for high-$z$ AGNs, where the source fully falls within the spectroscopic fiber or slit, as demonstrated by \cite{Moran2002} and later supported by \cite{Trump2009} for a sample of high-$z$ dull AGNs in the COSMOS survey. However, our sample is limited to $z<0.2,$ and furthermore, relevant studies showed that there is no significant dilution in dull local AGN samples \citep{LaFranca2002, Hornschemeier2005}. In addition, \citet{Rigby2006} argued against dilution also in high-$z$ AGNs. Therefore, we do not consider this scenario as a possible explanation for our results. A co-alignment of the torus with the disc cannot explain the deficit of Sy2 galaxies in edge-on systems either when compared with the control SF sample. 

Finally, we note that the smaller (higher) fraction of Sy2 (Sy1) with respect to SF galaxies at high b/a values strengthens the hypothesis of a host galaxy contribution to the obscuration of the BLR. Specifically, due to the fact that in face-on galaxies the gas and the dust of the disc does not intervene between the observer and the active nuclei, the only obscurer of the BLR is the torus, which apparently in some cases is not sufficient to hide the BLR, giving rise to the deficiency of type 2 in face-on galaxies \citep[e.g.][]{Lacy2007}. 

In the context of matching the samples of Sy1, Sy2, and control SF galaxies to stellar mass, the lack of 'hidden' Sy2 at small b/a values compared to the unmatched case can be explained, as we point out in Sect. \ref{stellar}, by the fact that the matching procedure cuts the most massive Sy2 and Sy1 galaxies, which are also the most abundant. Taking into account that more massive spiral galaxies contain more dust \citep[e.g.][]{Whitaker2017}, by excluding them from our samples we effectively exclude those host galaxies that can obscure both the AGN BLR and the NLR at extreme edge-on configurations. This result further supports the scenario of a host galaxy contribution to the obscuration of the AGN activity.

\section{Conclusions}
The main purpose of the current work was to test aspects of the simplest unification paradigm by searching for differences in the properties of the host galaxies of various AGN types. For our purposes, we used (a) the SDSS DR14 spectroscopic galaxy catalogue, selecting sub-samples of Sy1 and Sy2, LINER, Composite and SF galaxies, limited to $z<0.2,$ and (b) the results of the Galaxy Zoo project regarding the Hubble-type morphology of these galaxies.

Our main results are listed below:
\begin{enumerate}
\item We find statistically significant differences -quantified by a KS two-sample test- of the various types of AGN host-galaxy Hubble types, with the most significant result being that the fraction of Sy1 galaxies hosted by ellipticals is higher than that of any other AGN class. These results can be interpreted  within  a possible co-evolution scenario between galaxies and SMBS. 
\item  We also find that the orientation distributions, as revealed by the disc axis ratio ($b/a$) of the Sy1 and Sy2 spiral populations show statistically significant differences with respect to the control sample of star-forming galaxies (which by definition should cover all possible orientation configurations), which is in conflict with the predictions of the simple unified model. These differences hint towards an effect by which the dusty galactic disc has a significant contribution to the obscuration of the broad-line and partially also of the narrow-line nuclear region. This could also interpret our previous result regarding the host-galaxy Hubble types of type 1 AGN.
Indeed, the fact that we detect more Sy1 than any other AGN type in elliptical hosts (which as is well known are gas and dust deficient) can be explained if the amount of galactic dust and gas contributes to the obscuration of the nuclear region and in particular the BLR.
\end{enumerate}
These findings are at odds with the expectations of the simplest formulation of the unification paradigm, implying that beyond the orientation of the torus, at least two other factors play an important role in the AGN classification: (a) the orientation of the host galaxy, and (b) evolution.

\begin{acknowledgements}
Authors thank the anonymous referee for useful comments which helped us improve the presentation of this work.

      This paper has made use of the data from the SDSS projects. Funding for the Sloan Digital Sky Survey IV has been provided by the Alfred P. Sloan Foundation, the U.S. Department of
Energy Office of Science, and the Participating Institutions. SDSS-IV acknowledges support and resources from the Center for
High-Performance Computing at the University of Utah. The SDSS web site is www.sdss.org. 

SDSS-IV is managed by the Astrophysical Research Consortium for the Participating Institutions of the SDSS Collaboration including the Brazilian Participation
QUASAR BLACK HOLE MASSES 25
Group, the Carnegie Institution for Science, Carnegie Mellon University, the Chilean Participation Group, the French Participation Group, Harvard-Smithsonian Center for Astrophysics, Instituto de Astrofisica de Canarias, The Johns Hopkins University,
Kavli Institute for the Physics and Mathematics of the Universe (IPMU) / University of Tokyo, the Korean Participation Group,
Lawrence Berkeley National Laboratory, Leibniz Institut f\"ur Astrophysik Potsdam (AIP), Max-Planck-Institut f\"ur Astronomie
(MPIA Heidelberg), Max-Planck-Institut f\"ur Astrophysik (MPA Garching), Max-Planck-Institut f\"ur Extraterrestrische Physik
(MPE), National Astronomical Observatories of China, New Mexico State University, New York University, University of Notre
Dame, Observatario Nacional / MCTI, The Ohio State University, Pennsylvania State University, Shanghai Astronomical Observatory, United Kingdom Participation Group, Universidad Nacional Autonoma de Mexico, University of Arizona, University
of Colorado Boulder, University of Oxford, University of Portsmouth, University of Utah, University of Virginia, University of
Washington, University of Wisconsin, Vanderbilt University, and Yale University.
\end{acknowledgements}

\bibliographystyle{aa}
\bibliography{references}

\begin{thebibliography}{90}
\expandafter\ifx\csname natexlab\endcsname\relax\def\natexlab#1{#1}\fi

\bibitem[{{Abolfathi} {et~al.}(2018){Abolfathi}, {Aguado}, {Aguilar}, {Allende
  Prieto}, {Almeida}, {Ananna}, {Anders}, {Anderson}, {Andrews}, {Anguiano},
  {Arag{\'o}n-Salamanca}, {Argudo-Fern{\'a}ndez}, {Armengaud}, {Ata},
  {Aubourg}, {Avila-Reese}, {Badenes}, {Bailey}, {Balland}, {Barger},
  {Barrera-Ballesteros}, {Bartosz}, {Bastien}, {Bates}, {Baumgarten},
  {Bautista}, {Beaton}, {Beers}, {Belfiore}, {Bender}, {Bernardi}, {Bershady},
  {Beutler}, {Bird}, {Bizyaev}, {Blanc}, {Blanton}, {Blomqvist}, {Bolton},
  {Boquien}, {Borissova}, {Bovy}, {Bradna Diaz}, {Brandt}, {Brinkmann},
  {Brownstein}, {Bundy}, {Burgasser}, {Burtin}, {Busca}, {Ca{\~n}as},
  {Cano-D{\'\i}az}, {Cappellari}, {Carrera}, {Casey}, {Cervantes Sodi}, {Chen},
  {Cherinka}, {Chiappini}, {Choi}, {Chojnowski}, {Chuang}, {Chung}, {Clerc},
  {Cohen}, {Comerford}, {Comparat}, {Correa do Nascimento}, {da Costa},
  {Cousinou}, {Covey}, {Crane}, {Cruz-Gonzalez}, {Cunha}, {da Silva Ilha},
  {Damke}, {Darling}, {Davidson}, {Dawson}, {de Icaza Lizaola}, {de la
  Macorra}, {de la Torre}, {De Lee}, {de Sainte Agathe}, {Deconto Machado},
  {Dell'Agli}, {Delubac}, {Diamond-Stanic}, {Donor}, {Downes}, {Drory}, {du Mas
  des Bourboux}, {Duckworth}, {Dwelly}, {Dyer}, {Ebelke}, {Davis Eigenbrot},
  {Eisenstein}, {Elsworth}, {Emsellem}, {Eracleous}, {Erfanianfar},
  {Escoffier}, {Fan}, {Fern{\'a}ndez Alvar}, {Fernandez-Trincado}, {Fernand o
  Cirolini}, {Feuillet}, {Finoguenov}, {Fleming}, {Font-Ribera}, {Freischlad},
  {Frinchaboy}, {Fu}, {G{\'o}mez Maqueo Chew}, {Galbany}, {Garc{\'\i}a
  P{\'e}rez}, {Garcia-Dias}, {Garc{\'\i}a-Hern{\'a}ndez}, {Garma Oehmichen},
  {Gaulme}, {Gelfand }, {Gil-Mar{\'\i}n}, {Gillespie}, {Goddard}, {Gonz{\'a}lez
  Hern{\'a}ndez}, {Gonzalez-Perez}, {Grabowski}, {Green}, {Grier}, {Gueguen},
  {Guo}, {Guy}, {Hagen}, {Hall}, {Harding}, {Hasselquist}, {Hawley}, {Hayes},
  {Hearty}, {Hekker}, {Hernand ez}, {Hernandez Toledo}, {Hogg},
  {Holley-Bockelmann}, {Holtzman}, {Hou}, {Hsieh}, {Hunt}, {Hutchinson},
  {Hwang}, {Jimenez Angel}, {Johnson}, {Jones}, {J{\"o}nsson}, {Jullo}, {Khan},
  {Kinemuchi}, {Kirkby}, {Kirkpatrick}, {Kitaura}, {Knapp}, {Kneib},
  {Kollmeier}, {Lacerna}, {Lane}, {Lang}, {Law}, {Le Goff}, {Lee}, {Li}, {Li},
  {Lian}, {Liang}, {Lima}, {Lin}, {Long}, {Lucatello}, {Lundgren}, {Mackereth},
  {MacLeod}, {Mahadevan}, {Maia}, {Majewski}, {Manchado}, {Maraston},
  {Mariappan}, {Marques-Chaves}, {Masseron}, {Masters}, {McDermid}, {McGreer},
  {Melendez}, {Meneses-Goytia}, {Merloni}, {Merrifield}, {Meszaros}, {Meza},
  {Minchev}, {Minniti}, {Mueller}, {Muller-Sanchez}, {Muna}, {Mu{\~n}oz},
  {Myers}, {Nair}, {Nand ra}, {Ness}, {Newman}, {Nichol}, {Nidever},
  {Nitschelm}, {Noterdaeme}, {O'Connell}, {Oelkers}, {Oravetz}, {Oravetz},
  {Ort{\'\i}z}, {Osorio}, {Pace}, {Padilla}, {Palanque-Delabrouille},
  {Palicio}, {Pan}, {Pan}, {Parikh}, {P{\^a}ris}, {Park}, {Peirani},
  {Pellejero-Ibanez}, {Penny}, {Percival}, {Perez-Fournon}, {Petitjean},
  {Pieri}, {Pinsonneault}, {Pisani}, {Prada}, {Prakash}, {Queiroz}, {Raddick},
  {Raichoor}, {Barboza Rembold}, {Richstein}, {Riffel}, {Riffel}, {Rix},
  {Robin}, {Rodr{\'\i}guez Torres}, {Rom{\'a}n-Z{\'u}{\~n}iga}, {Ross},
  {Rossi}, {Ruan}, {Ruggeri}, {Ruiz}, {Salvato}, {S{\'a}nchez}, {S{\'a}nchez},
  {Sanchez Almeida}, {S{\'a}nchez-Gallego}, {Santana Rojas}, {Santiago},
  {Schiavon}, {Schimoia}, {Schlafly}, {Schlegel}, {Schneider}, {Schuster},
  {Schwope}, {Seo}, {Serenelli}, {Shen}, {Shen}, {Shetrone}, {Shull}, {Silva
  Aguirre}, {Simon}, {Skrutskie}, {Slosar}, {Smethurst}, {Smith}, {Sobeck},
  {Somers}, {Souter}, {Souto}, {Spindler}, {Stark}, {Stassun}, {Steinmetz},
  {Stello}, {Storchi-Bergmann}, {Streblyanska}, {Stringfellow}, {Su{\'a}rez},
  {Sun}, {Szigeti}, {Taghizadeh-Popp}, {Talbot}, {Tang}, {Tao}, {Tayar},
  {Tembe}, {Teske}, {Thakar}, {Thomas}, {Tissera}, {Tojeiro}, {Tremonti},
  {Troup}, {Urry}, {Valenzuela}, {van den Bosch}, {Vargas-Gonz{\'a}lez},
  {Vargas-Maga{\~n}a}, {Vazquez}, {Villanova}, {Vogt}, {Wake}, {Wang},
  {Weaver}, {Weijmans}, {Weinberg}, {Westfall}, {Whelan}, {Wilcots}, {Wild},
  {Williams}, {Wilson}, {Wood-Vasey}, {Wylezalek}, {Xiao}, {Yan}, {Yang},
  {Ybarra}, {Y{\`e}che}, {Zakamska}, {Zamora}, {Zarrouk}, {Zasowski}, {Zhang},
  {Zhao}, {Zhao}, {Zheng}, {Zheng}, {Zhou}, {Zhu}, {Zinn}, \&
  {Zou}}]{Abolfathi2018}
{Abolfathi}, B., {Aguado}, D.~S., {Aguilar}, G., {et~al.} 2018, \apjs, 235, 42

\bibitem[{{Antonucci}(1993)}]{Antonucci1993}
{Antonucci}, R. 1993, \araa, 31, 473

\bibitem[{{Audibert} {et~al.}(2017){Audibert}, {Riffel}, {Sales}, {Pastoriza},
  \& {Ruschel-Dutra}}]{Audibert2017}
{Audibert}, A., {Riffel}, R., {Sales}, D.~A., {Pastoriza}, M.~G., \&
  {Ruschel-Dutra}, D. 2017, \mnras, 464, 2139

\bibitem[{{Baldwin} {et~al.}(1981){Baldwin}, {Phillips}, \&
  {Terlevich}}]{Baldwin1981}
{Baldwin}, J.~A., {Phillips}, M.~M., \& {Terlevich}, R. 1981, \pasp, 93, 5

\bibitem[{{Bitsakis} {et~al.}(2010){Bitsakis}, {Charmandaris}, {Le Floc'h},
  {D{\'\i}az-Santos}, {Slater}, {Xilouris}, \& {Haynes}}]{Bitsakis2010}
{Bitsakis}, T., {Charmandaris}, V., {Le Floc'h}, E., {et~al.} 2010, \aap, 517,
  A75

\bibitem[{{Bornancini} \& {Garc{\'\i}a Lambas}(2018)}]{Bornancini2018}
{Bornancini}, C. \& {Garc{\'\i}a Lambas}, D. 2018, \mnras, 479, 2308

\bibitem[{{Bornancini} \& {Garc{\'\i}a Lambas}(2020)}]{Bornacini2020}
{Bornancini}, C. \& {Garc{\'\i}a Lambas}, D. 2020, \mnras, 494, 1189

\bibitem[{{Bornancini} {et~al.}(2017){Bornancini}, {Taormina}, \&
  {Lambas}}]{Bornancini2017}
{Bornancini}, C.~G., {Taormina}, M.~S., \& {Lambas}, D.~G. 2017, \aap, 605, A10

\bibitem[{{Brinchmann} {et~al.}(2004){Brinchmann}, {Charlot}, {Heckman},
  {Kauffmann}, {Tremonti}, \& {White}}]{Brinchmann2004}
{Brinchmann}, J., {Charlot}, S., {Heckman}, T.~M., {et~al.} 2004, arXiv
  e-prints, astro

\bibitem[{{Buchner} \& {Bauer}(2017)}]{Buchner2017}
{Buchner}, J. \& {Bauer}, F.~E. 2017, \mnras, 465, 4348

\bibitem[{{Burtscher} {et~al.}(2016){Burtscher}, {Davies}, {Graci{\'a}-Carpio},
  {Koss}, {Lin}, {Lutz}, {Nandra}, {Netzer}, {Orban de Xivry}, {Ricci},
  {Rosario}, {Veilleux}, {Contursi}, {Genzel}, {Schnorr-M{\"u}ller},
  {Sternberg}, {Sturm}, \& {Tacconi}}]{Burtscher2016}
{Burtscher}, L., {Davies}, R.~I., {Graci{\'a}-Carpio}, J., {et~al.} 2016, \aap,
  586, A28

\bibitem[{{Chen} {et~al.}(2015){Chen}, {Hickox}, {Alberts}, {Harrison},
  {Alexander}, {Assef}, {Brodwin}, {Brown}, {Del Moro}, {Forman}, {Gorjian},
  {Goulding}, {Hainline}, {Jones}, {Kochanek}, {Murray}, {Pope}, {Rovilos}, \&
  {Stern}}]{Chen2015}
{Chen}, C.-T.~J., {Hickox}, R.~C., {Alberts}, S., {et~al.} 2015, \apj, 802, 50

\bibitem[{Chen \& Hwang(2017)}]{Chen2017}
Chen, Y.-C. \& Hwang, C.-Y. 2017, Astrophysics and Space Science, 362

\bibitem[{{DiPompeo} {et~al.}(2017{\natexlab{a}}){DiPompeo}, {Hickox},
  {Eftekharzadeh}, \& {Myers}}]{DiPompeo2017a}
{DiPompeo}, M.~A., {Hickox}, R.~C., {Eftekharzadeh}, S., \& {Myers}, A.~D.
  2017{\natexlab{a}}, \mnras, 469, 4630

\bibitem[{{DiPompeo} {et~al.}(2017{\natexlab{b}}){DiPompeo}, {Hickox}, {Myers},
  \& {Geach}}]{DiPompeo2017b}
{DiPompeo}, M.~A., {Hickox}, R.~C., {Myers}, A.~D., \& {Geach}, J.~E.
  2017{\natexlab{b}}, \mnras, 464, 3526

\bibitem[{{Elitzur}(2012)}]{Elitzur2012}
{Elitzur}, M. 2012, \apjl, 747, L33

\bibitem[{{Elitzur} \& {Ho}(2009)}]{ElitzurHo2009}
{Elitzur}, M. \& {Ho}, L.~C. 2009, \apjl, 701, L91

\bibitem[{{Elitzur} {et~al.}(2014){Elitzur}, {Ho}, \&
  {Trump}}]{ElitzurTrump2014}
{Elitzur}, M., {Ho}, L.~C., \& {Trump}, J.~R. 2014, \mnras, 438, 3340

\bibitem[{{Elitzur} \& {Shlosman}(2006)}]{Elitzur2006}
{Elitzur}, M. \& {Shlosman}, I. 2006, \apjl, 648, L101

\bibitem[{{Ferrarese} \& {Merritt}(2000)}]{Ferrarese2000}
{Ferrarese}, L. \& {Merritt}, D. 2000, \apjl, 539, L9

\bibitem[{{Gebhardt} {et~al.}(2000){Gebhardt}, {Bender}, {Bower}, {Dressler},
  {Faber}, {Filippenko}, {Green}, {Grillmair}, {Ho}, {Kormendy}, {Lauer},
  {Magorrian}, {Pinkney}, {Richstone}, \& {Tremaine}}]{Gebhardt2000}
{Gebhardt}, K., {Bender}, R., {Bower}, G., {et~al.} 2000, \apjl, 539, L13

\bibitem[{{Gonz{\'a}lez} {et~al.}(2008){Gonz{\'a}lez}, {Krongold}, {Dultzin},
  {Hern{\'a}ndez-Toledo}, {Huerta}, {Olgu{\'\i}n}, {Marziani}, \&
  {Cruz-Gonz{\'a}lez}}]{Gonzales2008}
{Gonz{\'a}lez}, J.~J., {Krongold}, Y., {Dultzin}, D., {et~al.} 2008, in Revista
  Mexicana de Astronomia y Astrofisica Conference Series, Vol.~32, Revista
  Mexicana de Astronomia y Astrofisica Conference Series, 170--172

\bibitem[{{Goulding} {et~al.}(2012){Goulding}, {Alexander}, {Bauer}, {Forman},
  {Hickox}, {Jones}, {Mullaney}, \& {Trichas}}]{Goulding2012}
{Goulding}, A.~D., {Alexander}, D.~M., {Bauer}, F.~E., {et~al.} 2012, \apj,
  755, 5

\bibitem[{{He} {et~al.}(2018){He}, {Sun}, {Zakamska}, {Wylezalek}, {Kelly},
  {Greene}, {Rembold}, {Riffel}, \& {Riffel}}]{Zhicheng2018}
{He}, Z., {Sun}, A.-L., {Zakamska}, N.~L., {et~al.} 2018, \mnras, 478, 3614

\bibitem[{{Heckman} \& {Best}(2014)}]{Heckman2014}
{Heckman}, T.~M. \& {Best}, P.~N. 2014, \araa, 52, 589

\bibitem[{{Hern{\'a}ndez-Ibarra} {et~al.}(2016){Hern{\'a}ndez-Ibarra},
  {Krongold}, {Dultzin}, {del Olmo}, {Perea}, {Gonz{\'a}lez},
  {Mendoza-Castrej{\'o}n}, \& {Bitsakis}}]{Hernandez2016}
{Hern{\'a}ndez-Ibarra}, F.~J., {Krongold}, Y., {Dultzin}, D., {et~al.} 2016,
  \mnras, 459, 291

\bibitem[{{Hickox} \& {Alexander}(2018)}]{Hickox2018}
{Hickox}, R.~C. \& {Alexander}, D.~M. 2018, \araa, 56, 625

\bibitem[{{Hopkins} {et~al.}(2008{\natexlab{a}}){Hopkins}, {Cox},
  {Kere{\v{s}}}, \& {Hernquist}}]{Hopkins2008b}
{Hopkins}, P.~F., {Cox}, T.~J., {Kere{\v{s}}}, D., \& {Hernquist}, L.
  2008{\natexlab{a}}, \apjs, 175, 390

\bibitem[{{Hopkins} {et~al.}(2006){Hopkins}, {Hernquist}, {Cox}, {Di Matteo},
  {Robertson}, \& {Springel}}]{Hopkins2006}
{Hopkins}, P.~F., {Hernquist}, L., {Cox}, T.~J., {et~al.} 2006, \apjs, 163, 1

\bibitem[{{Hopkins} {et~al.}(2008{\natexlab{b}}){Hopkins}, {Hernquist}, {Cox},
  \& {Kere{\v{s}}}}]{Hopkins2008a}
{Hopkins}, P.~F., {Hernquist}, L., {Cox}, T.~J., \& {Kere{\v{s}}}, D.
  2008{\natexlab{b}}, \apjs, 175, 356

\bibitem[{{Hornschemeier} {et~al.}(2005){Hornschemeier}, {Heckman}, {Ptak},
  {Tremonti}, \& {Colbert}}]{Hornschemeier2005}
{Hornschemeier}, A.~E., {Heckman}, T.~M., {Ptak}, A.~F., {Tremonti}, C.~A., \&
  {Colbert}, E.~J.~M. 2005, \aj, 129, 86

\bibitem[{{Hubble}(1926)}]{Hubble1926}
{Hubble}, E. 1926, Contributions from the Mount Wilson Observatory / Carnegie
  Institution of Washington, 324, 1

\bibitem[{{Jiang} {et~al.}(2016){Jiang}, {Wang}, {Mo}, {Dong}, {Wang}, \&
  {Zhou}}]{Jiang2016}
{Jiang}, N., {Wang}, H., {Mo}, H., {et~al.} 2016, \apj, 832, 111

\bibitem[{{Kauffmann} {et~al.}(2003){Kauffmann}, {Heckman}, {White}, {Charlot},
  {Tremonti}, {Peng}, {Seibert}, {Brinkmann}, {Nichol}, {SubbaRao}, \&
  {York}}]{Kauffmann2003}
{Kauffmann}, G., {Heckman}, T.~M., {White}, S. D.~M., {et~al.} 2003, \mnras,
  341, 54

\bibitem[{{Keel}(1980)}]{Keel1980}
{Keel}, W.~C. 1980, \aj, 85, 198

\bibitem[{Kinney {et~al.}(2000)Kinney, Schmitt, Clarke, Pringle, Ulvestad, \&
  Antonucci}]{Kinney2000}
Kinney, A.~L., Schmitt, H.~R., Clarke, C.~J., {et~al.} 2000, The Astrophysical
  Journal, 537, 152–177

\bibitem[{{Kormendy} \& {Ho}(2013)}]{Kormendy2013}
{Kormendy}, J. \& {Ho}, L.~C. 2013, \araa, 51, 511

\bibitem[{{Koulouridis}(2014)}]{Koulouridis2014}
{Koulouridis}, E. 2014, \aap, 570, A72

\bibitem[{{Koulouridis} {et~al.}(2013){Koulouridis}, {Plionis}, {Chavushyan},
  {Dultzin}, {Krongold}, {Georgantopoulos}, \&
  {Le{\'o}n-Tavares}}]{Koulouridis2013}
{Koulouridis}, E., {Plionis}, M., {Chavushyan}, V., {et~al.} 2013, \aap, 552,
  A135

\bibitem[{{Koulouridis} {et~al.}(2006){Koulouridis}, {Plionis}, {Chavushyan},
  {Dultzin-Hacyan}, {Krongold}, \& {Goudis}}]{Koulouridis2006}
{Koulouridis}, E., {Plionis}, M., {Chavushyan}, V., {et~al.} 2006, \apj, 639,
  37

\bibitem[{{Krongold} {et~al.}(2002){Krongold}, {Dultzin-Hacyan}, \&
  {Marziani}}]{Krongold2002}
{Krongold}, Y., {Dultzin-Hacyan}, D., \& {Marziani}, P. 2002, \apj, 572, 169

\bibitem[{{La Franca} {et~al.}(2002){La Franca}, {Fiore}, {Vignali},
  {Antonelli}, {Comastri}, {Giommi}, {Matt}, {Molendi}, {Perola}, \&
  {Pompilio}}]{LaFranca2002}
{La Franca}, F., {Fiore}, F., {Vignali}, C., {et~al.} 2002, \apj, 570, 100

\bibitem[{{Lacy} {et~al.}(2007){Lacy}, {Sajina}, {Petric}, {Seymour},
  {Canalizo}, {Ridgway}, {Armus}, \& {Storrie-Lombardi}}]{Lacy2007}
{Lacy}, M., {Sajina}, A., {Petric}, A.~O., {et~al.} 2007, \apjl, 669, L61

\bibitem[{{Lagos} {et~al.}(2011){Lagos}, {Padilla}, {Strauss}, {Cora}, \&
  {Hao}}]{Lagos2011}
{Lagos}, C. D.~P., {Padilla}, N.~D., {Strauss}, M.~A., {Cora}, S.~A., \& {Hao},
  L. 2011, \mnras, 414, 2148

\bibitem[{{Lanzuisi} {et~al.}(2017){Lanzuisi}, {Delvecchio}, {Berta}, {Brusa},
  {Comastri}, {Gilli}, {Gruppioni}, {Marchesi}, {Perna}, {Pozzi}, {Salvato},
  {Symeonidis}, {Vignali}, {Vito}, {Volonteri}, \& {Zamorani}}]{Lanzuisi2017}
{Lanzuisi}, G., {Delvecchio}, I., {Berta}, S., {et~al.} 2017, \aap, 602, A123

\bibitem[{{Lawrence} \& {Elvis}(1982)}]{Lawrence1982}
{Lawrence}, A. \& {Elvis}, M. 1982, \apj, 256, 410

\bibitem[{{Lintott} {et~al.}(2008){Lintott}, {Schawinski}, {Slosar}, {Land},
  {Bamford}, {Thomas}, {Raddick}, {Nichol}, {Szalay}, \&
  {Andreescu}}]{Lintott2008}
{Lintott}, C.~J., {Schawinski}, K., {Slosar}, A., {et~al.} 2008, \mnras, 389,
  1179

\bibitem[{Madau \& Dickinson(2014)}]{Madau2014}
Madau, P. \& Dickinson, M. 2014, Annual Review of Astronomy and Astrophysics,
  52, 415–486

\bibitem[{{Magorrian} {et~al.}(1998){Magorrian}, {Tremaine}, {Richstone},
  {Bender}, {Bower}, {Dressler}, {Faber}, {Gebhardt}, {Green}, {Grillmair},
  {Kormendy}, \& {Lauer}}]{Magorrian1998}
{Magorrian}, J., {Tremaine}, S., {Richstone}, D., {et~al.} 1998, \aj, 115, 2285

\bibitem[{{Maiolino} \& {Rieke}(1995)}]{Maiolino1995}
{Maiolino}, R. \& {Rieke}, G.~H. 1995, \apj, 454, 95

\bibitem[{{Malizia} {et~al.}(2020){Malizia}, {Bassani}, {Stephen}, {Bazzano},
  \& {Ubertini}}]{Malizia2020}
{Malizia}, A., {Bassani}, L., {Stephen}, J.~B., {Bazzano}, A., \& {Ubertini},
  P. 2020, \aap, 639, A5

\bibitem[{{Malkan} {et~al.}(1998){Malkan}, {Gorjian}, \& {Tam}}]{Malkan1998}
{Malkan}, M.~A., {Gorjian}, V., \& {Tam}, R. 1998, \apjs, 117, 25

\bibitem[{{Mart{\'\i}nez} {et~al.}(2008){Mart{\'\i}nez}, {del Olmo}, {Coziol},
  \& {Focardi}}]{Martinez2008}
{Mart{\'\i}nez}, M.~A., {del Olmo}, A., {Coziol}, R., \& {Focardi}, P. 2008,
  \apjl, 678, L9

\bibitem[{{Mart{\'\i}nez-Sansigre}
  {et~al.}(2006{\natexlab{a}}){Mart{\'\i}nez-Sansigre}, {Rawlings}, {Lacy},
  {Fadda}, {Marleau}, {Simpson}, {Willott}, \& {Jarvis}}]{Martinez2006}
{Mart{\'\i}nez-Sansigre}, A., {Rawlings}, S., {Lacy}, M., {et~al.}
  2006{\natexlab{a}}, Astronomische Nachrichten, 327, 266

\bibitem[{{Mart{\'\i}nez-Sansigre}
  {et~al.}(2006{\natexlab{b}}){Mart{\'\i}nez-Sansigre}, {Rawlings}, {Lacy},
  {Fadda}, {Marleau}, {Simpson}, {Willott}, \& {Jarvis}}]{Mart2006}
{Mart{\'\i}nez-Sansigre}, A., {Rawlings}, S., {Lacy}, M., {et~al.}
  2006{\natexlab{b}}, Astronomische Nachrichten, 327, 266

\bibitem[{{Masoura} {et~al.}(2021){Masoura}, {Mountrichas}, {Georgantopoulos},
  \& {Plionis}}]{Masoura2021}
{Masoura}, V.~A., {Mountrichas}, G., {Georgantopoulos}, I., \& {Plionis}, M.
  2021, arXiv e-prints, arXiv:2101.00724

\bibitem[{{Matt}(2000)}]{Matt2000}
{Matt}, G. 2000, \aap, 355, L31

\bibitem[{{McLeod} \& {Rieke}(1995)}]{McLeod1995}
{McLeod}, K.~K. \& {Rieke}, G.~H. 1995, \apj, 441, 96

\bibitem[{{Melnyk} {et~al.}(2018){Melnyk}, {Elyiv}, {Smol{\v{c}}i{\'c}},
  {Plionis}, {Koulouridis}, {Fotopoulou}, {Chiappetti}, {Adami}, {Baran},
  {Butler}, {Delhaize}, {Delvecchio}, {Finet}, {Huynh}, {Lidman}, {Pierre},
  {Pompei}, {Vignali}, \& {Surdej}}]{Melnyk2018}
{Melnyk}, O., {Elyiv}, A., {Smol{\v{c}}i{\'c}}, V., {et~al.} 2018, \aap, 620,
  A6

\bibitem[{{Mendoza-Castrej{\'o}n} {et~al.}(2015){Mendoza-Castrej{\'o}n},
  {Dultzin}, {Krongold}, {Gonz{\'a}lez}, \& {Elitzur}}]{Mendoza2015}
{Mendoza-Castrej{\'o}n}, S., {Dultzin}, D., {Krongold}, Y., {Gonz{\'a}lez},
  J.~J., \& {Elitzur}, M. 2015, \mnras, 447, 2437

\bibitem[{{Moran} \& {Filippenko}(2002)}]{Moran2002}
{Moran}, E.~C. \& {Filippenko}, A.~V. 2002, in American Astronomical Society
  Meeting Abstracts, Vol. 200, American Astronomical Society Meeting Abstracts
  \#200, 17.04

\bibitem[{{Netzer}(2015)}]{Netzer2015}
{Netzer}, H. 2015, \araa, 53, 365

\bibitem[{{Nicastro} {et~al.}(2003){Nicastro}, {Martocchia}, \&
  {Matt}}]{Nicastro2003}
{Nicastro}, F., {Martocchia}, A., \& {Matt}, G. 2003, \apjl, 589, L13

\bibitem[{{Perlman} {et~al.}(2007){Perlman}, {Mason}, {Packham}, {Levenson},
  {Elitzur}, {Schaefer}, {Imanishi}, {Sparks}, \& {Radomski}}]{Perlman2007}
{Perlman}, E.~S., {Mason}, R.~E., {Packham}, C., {et~al.} 2007, \apj, 663, 808

\bibitem[{Powell {et~al.}(2018)Powell, Cappelluti, Urry, Koss, Finoguenov,
  Ricci, Trakhtenbrot, Allevato, Ajello, Oh, \& et~al.}]{Powell2018}
Powell, M.~C., Cappelluti, N., Urry, C.~M., {et~al.} 2018, The Astrophysical
  Journal, 858, 110

\bibitem[{{Prieto} {et~al.}(2014){Prieto}, {Mezcua}, {Fern{\'a}ndez-Ontiveros},
  \& {Schartmann}}]{Prieto2014}
{Prieto}, M.~A., {Mezcua}, M., {Fern{\'a}ndez-Ontiveros}, J.~A., \&
  {Schartmann}, M. 2014, \mnras, 442, 2145

\bibitem[{{Raddick} {et~al.}(2007){Raddick}, {Lintott}, {Schawinski}, {Thomas},
  {Nichol}, {Andreescu}, {Bamford}, {Land}, {Murray}, {Slosar}, {Szalay},
  {Vandenberg}, \& {Galaxy Zoo Team}}]{Raddick2007}
{Raddick}, J., {Lintott}, C.~J., {Schawinski}, K., {et~al.} 2007, in American
  Astronomical Society Meeting Abstracts, Vol. 211, 94.03

\bibitem[{Ramos~Almeida {et~al.}(2011)Ramos~Almeida, Levenson, Alonso-Herrero,
  Asensio~Ramos, Rodríguez~Espinosa, Pérez~García, Packham, Mason, Radomski,
  \& Díaz-Santos}]{Almeida2011}
Ramos~Almeida, C., Levenson, N.~A., Alonso-Herrero, A., {et~al.} 2011, The
  Astrophysical Journal, 731, 92

\bibitem[{{Rigby} {et~al.}(2006){Rigby}, {Rieke}, {Donley}, {Alonso-Herrero},
  \& {P{\'e}rez-Gonz{\'a}lez}}]{Rigby2006}
{Rigby}, J.~R., {Rieke}, G.~H., {Donley}, J.~L., {Alonso-Herrero}, A., \&
  {P{\'e}rez-Gonz{\'a}lez}, P.~G. 2006, \apj, 645, 115

\bibitem[{{Salim} {et~al.}(2007){Salim}, {Rich}, {Charlot}, {Brinchmann},
  {Johnson}, {Schiminovich}, {Seibert}, {Mallery}, {Heckman}, {Forster},
  {Friedman}, {Martin}, {Morrissey}, {Neff}, {Small}, {Wyder}, {Bianchi},
  {Donas}, {Lee}, {Madore}, {Milliard}, {Szalay}, {Welsh}, \& {Yi}}]{Salim2007}
{Salim}, S., {Rich}, R.~M., {Charlot}, S., {et~al.} 2007, \apjs, 173, 267

\bibitem[{{Schmitt} {et~al.}(2001){Schmitt}, {Antonucci}, {Ulvestad}, {Kinney},
  {Clarke}, \& {Pringle}}]{Schmitt2001}
{Schmitt}, H.~R., {Antonucci}, R.~R.~J., {Ulvestad}, J.~S., {et~al.} 2001,
  \apj, 555, 663

\bibitem[{{Siebenmorgen} {et~al.}(2015){Siebenmorgen}, {Heymann}, \&
  {Efstathiou}}]{Siebenmorgen2015}
{Siebenmorgen}, R., {Heymann}, F., \& {Efstathiou}, A. 2015, \aap, 583, A120

\bibitem[{{Slavcheva-Mihova} \& {Mihov}(2011)}]{Slavcheva2011}
{Slavcheva-Mihova}, L. \& {Mihov}, B. 2011, \aap, 526, A43

\bibitem[{{Sorrentino} {et~al.}(2006){Sorrentino}, {Radovich}, \&
  {Rifatto}}]{Sorrentino2006A}
{Sorrentino}, G., {Radovich}, M., \& {Rifatto}, A. 2006, \aap, 451, 809

\bibitem[{{Spinoglio} \& {Fern{\'a}ndez-Ontiveros}(2019)}]{Spinoglio2019}
{Spinoglio}, L. \& {Fern{\'a}ndez-Ontiveros}, J.~A. 2019, arXiv e-prints,
  arXiv:1911.12176

\bibitem[{{Springel} {et~al.}(2005){Springel}, {Di Matteo}, \&
  {Hernquist}}]{Springel2005}
{Springel}, V., {Di Matteo}, T., \& {Hernquist}, L. 2005, \mnras, 361, 776

\bibitem[{Suh {et~al.}(2019)Suh, Civano, Hasinger, Lusso, Marchesi, Schulze,
  Onodera, Rosario, \& Sanders}]{Suh2019}
Suh, H., Civano, F., Hasinger, G., {et~al.} 2019, The Astrophysical Journal,
  872, 168

\bibitem[{{Tran}(2001)}]{Tran2001T}
{Tran}, H.~D. 2001, \apjl, 554, L19

\bibitem[{{Tran}(2003)}]{Tran2003}
{Tran}, H.~D. 2003, \apj, 583, 632

\bibitem[{{Tremonti} {et~al.}(2004){Tremonti}, {Heckman}, {Kauffmann},
  {Brinchmann}, {Charlot}, {White}, {Seibert}, {Peng}, {Schlegel}, {Uomoto},
  {Fukugita}, \& {Brinkmann}}]{Tremonti2004}
{Tremonti}, C.~A., {Heckman}, T.~M., {Kauffmann}, G., {et~al.} 2004, \apj, 613,
  898

\bibitem[{{Trump} {et~al.}(2011){Trump}, {Impey}, \& {Kelly}}]{Trump2011}
{Trump}, J.~R., {Impey}, C.~D., \& {Kelly}, B.~C. 2011, in American
  Astronomical Society Meeting Abstracts, Vol. 217, American Astronomical
  Society Meeting Abstracts \#217, 430.08

\bibitem[{{Trump} {et~al.}(2009){Trump}, {Impey}, {Taniguchi}, {Brusa},
  {Civano}, {Elvis}, {Gabor}, {Jahnke}, {Kelly}, {Koekemoer}, {Nagao},
  {Salvato}, {Shioya}, {Capak}, {Huchra}, {Kartaltepe}, {Lanzuisi}, {McCarthy},
  {Maineri}, \& {Scoville}}]{Trump2009}
{Trump}, J.~R., {Impey}, C.~D., {Taniguchi}, Y., {et~al.} 2009, \apj, 706, 797

\bibitem[{{Urry} \& {Padovani}(1995)}]{Urrya1995}
{Urry}, C.~M. \& {Padovani}, P. 1995, \pasp, 107, 803

\bibitem[{{Villarroel} {et~al.}(2012){Villarroel}, {Korn}, \&
  {Matsuoka}}]{Villarroel2012}
{Villarroel}, B., {Korn}, A., \& {Matsuoka}, Y. 2012, arXiv e-prints,
  arXiv:1211.0528

\bibitem[{{Villarroel} \& {Korn}(2014)}]{Villarroel2014}
{Villarroel}, B. \& {Korn}, A.~J. 2014, Nature Physics, 10, 417

\bibitem[{{Villarroel} {et~al.}(2017){Villarroel}, {Nyholm}, {Karlsson},
  {Comer{\'o}n}, {Korn}, {Sollerman}, \& {Zackrisson}}]{Villarroel2017}
{Villarroel}, B., {Nyholm}, A., {Karlsson}, T., {et~al.} 2017, \apj, 837, 110

\bibitem[{{Whitaker} {et~al.}(2017){Whitaker}, {Pope}, {Cybulski}, {Casey},
  {Popping}, \& {Yun}}]{Whitaker2017}
{Whitaker}, K.~E., {Pope}, A., {Cybulski}, R., {et~al.} 2017, \apj, 850, 208

\bibitem[{{Yang} {et~al.}(2019){Yang}, {Brandt}, {Alexander}, {Chen}, {Ni},
  {Vito}, \& {Zhu}}]{Yang2019}
{Yang}, G., {Brandt}, W.~N., {Alexander}, D.~M., {et~al.} 2019, \mnras, 485,
  3721

\bibitem[{Yang {et~al.}(2017)}]{Yang2017}
Yang, G. {et~al.} 2017, Astrophys. J., 842, 72

\bibitem[{{Zou} {et~al.}(2019){Zou}, {Yang}, {Brandt}, \& {Xue}}]{Zou2019}
{Zou}, F., {Yang}, G., {Brandt}, W.~N., \& {Xue}, Y. 2019, arXiv e-prints,
  arXiv:1904.13286

\end{thebibliography}

\begin{appendix}

\section{Example of Sy1 and Sy2 spectra}

\begin{figure*} [h]
    \centering
    \begin{subfigure}[b]{0.32\textwidth}
        \includegraphics[width=\textwidth]{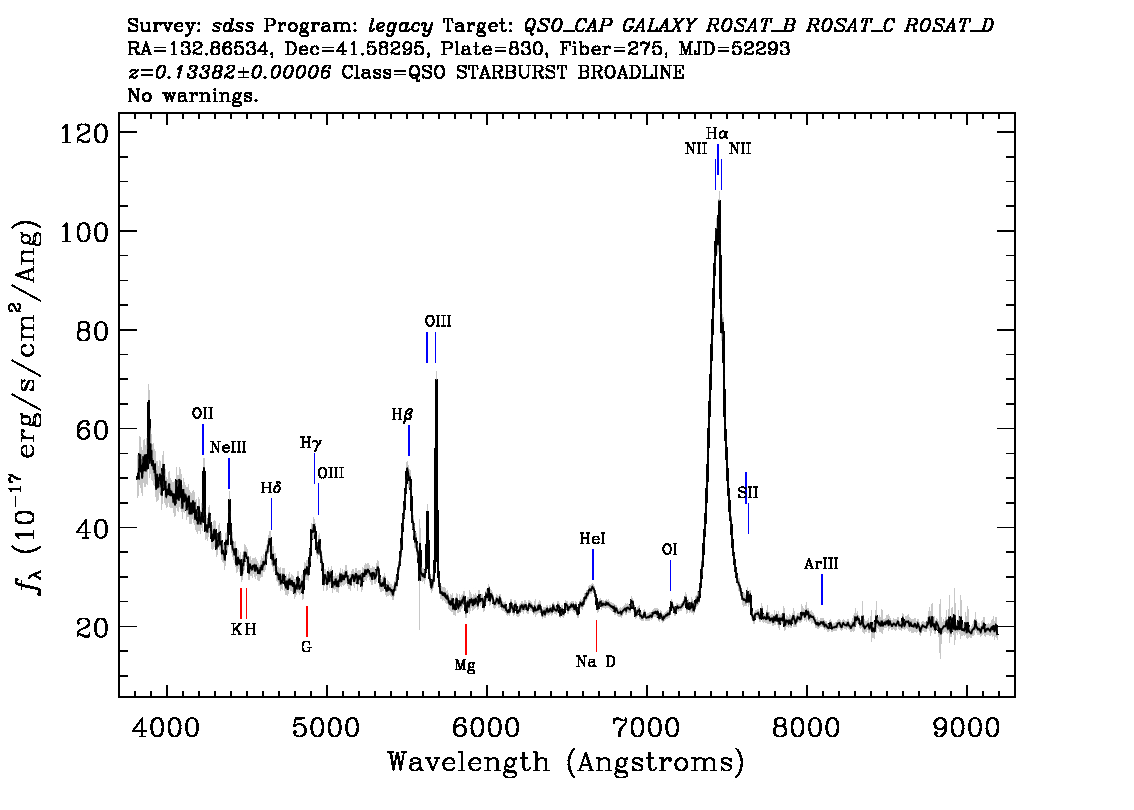}
    \end{subfigure}
    ~ 
    \begin{subfigure}[b]{0.32\textwidth}
      \includegraphics[width=\textwidth]{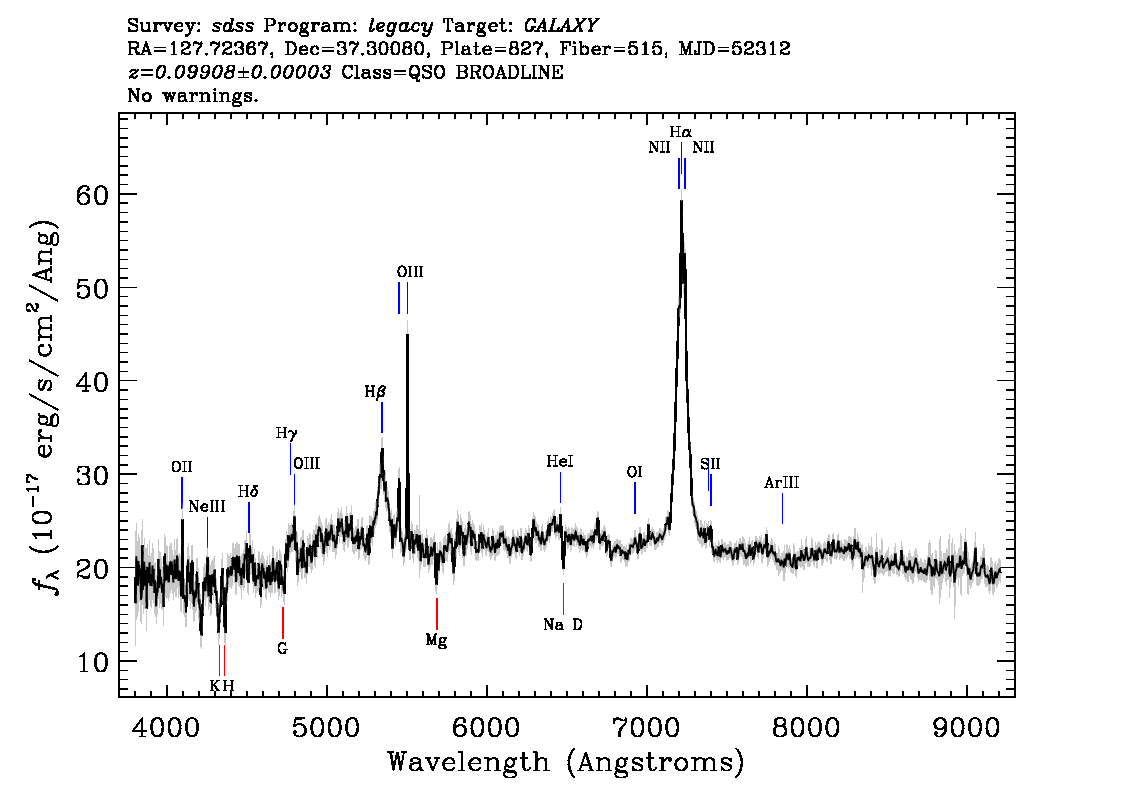}
    \end{subfigure}
    ~ 
    \begin{subfigure}[b]{0.32\textwidth}
        \includegraphics[width=\textwidth]{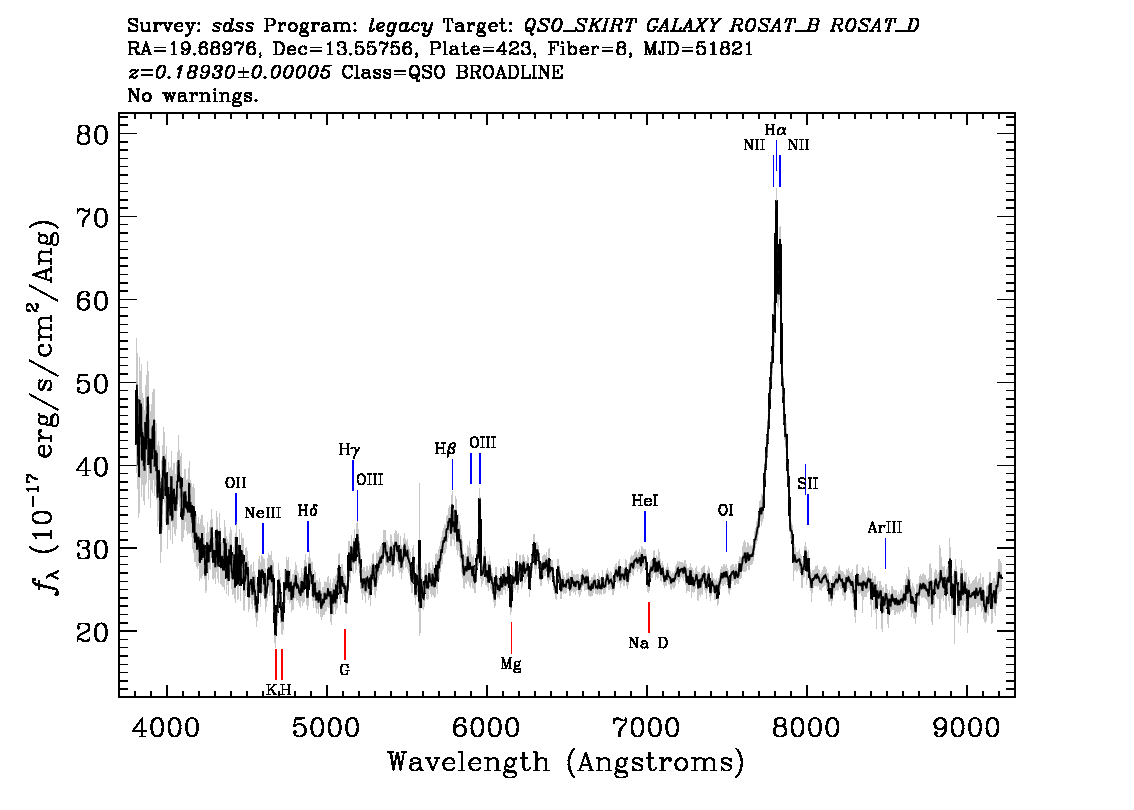}
    \end{subfigure}
    \begin{subfigure}[b]{0.32\textwidth}
        \includegraphics[width=\textwidth]{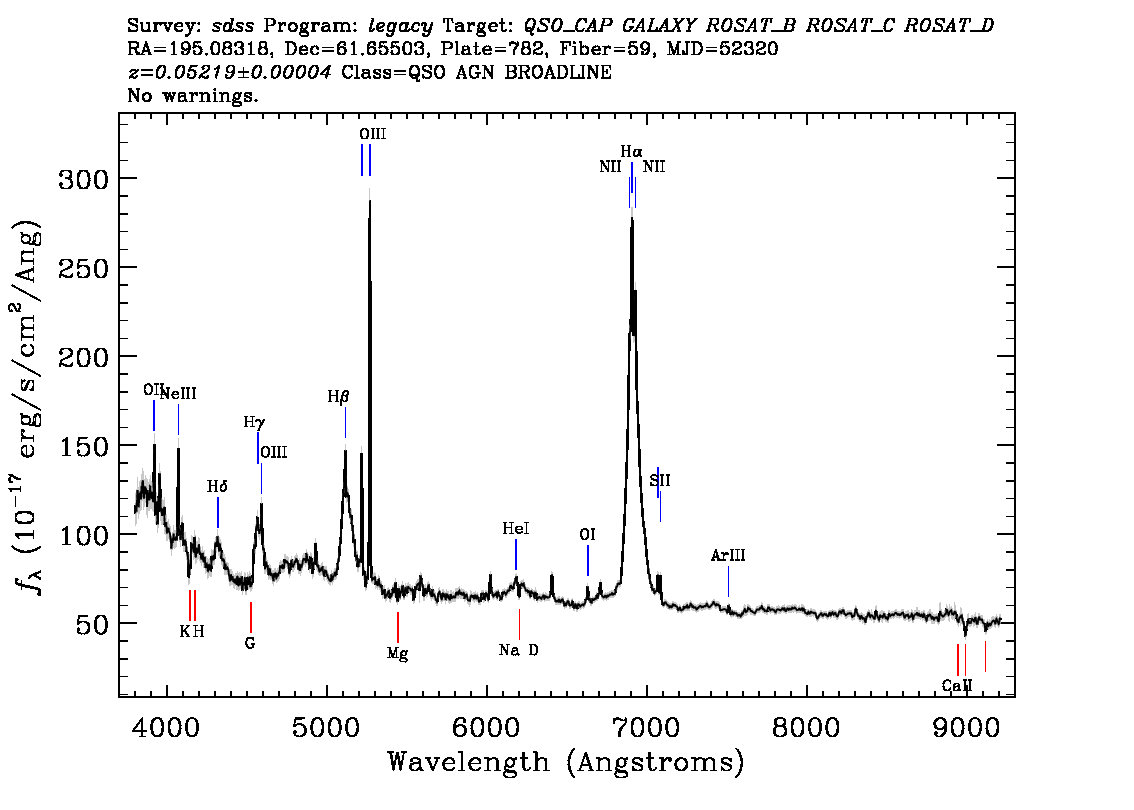}
    \end{subfigure}
    \begin{subfigure}[b]{0.32\textwidth}
        \includegraphics[width=\textwidth]{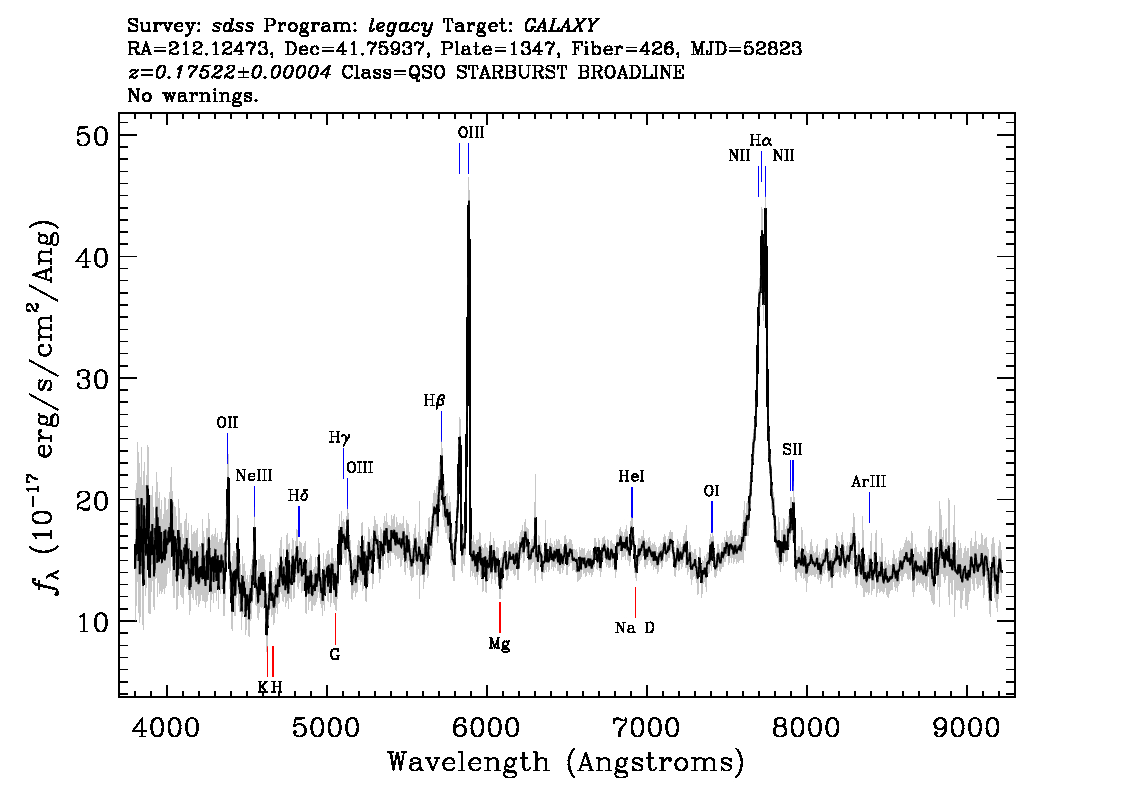}
    \end{subfigure}
    \begin{subfigure}[b]{0.32\textwidth}
        \includegraphics[width=\textwidth]{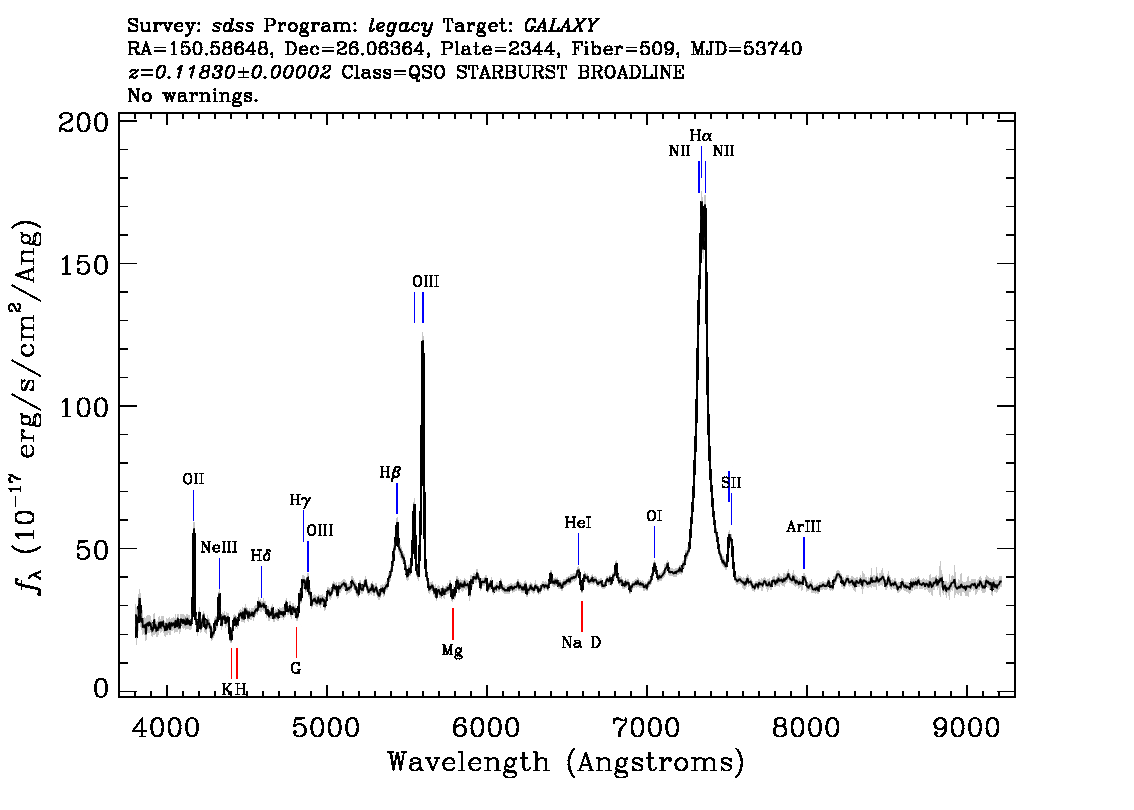}
    \end{subfigure}
    \begin{subfigure}[b]{0.32\textwidth}
        \includegraphics[width=\textwidth]{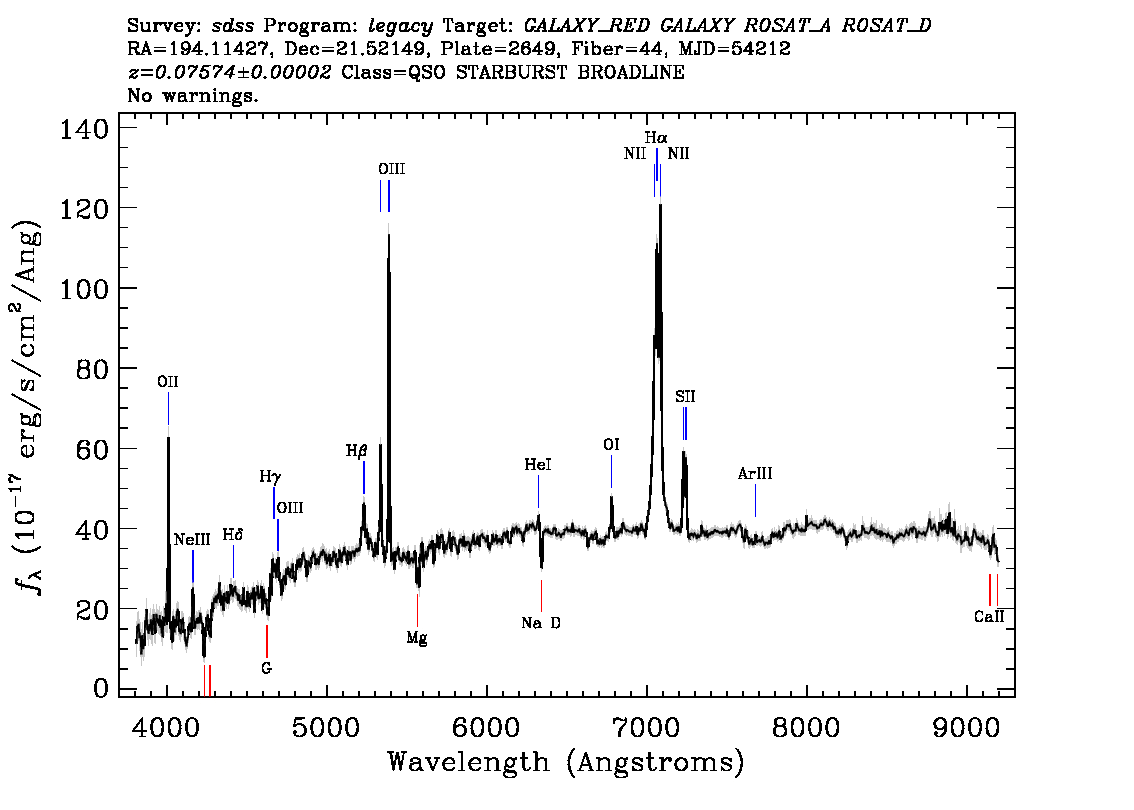}
    \end{subfigure}
    \begin{subfigure}[b]{0.32\textwidth}
        \includegraphics[width=\textwidth]{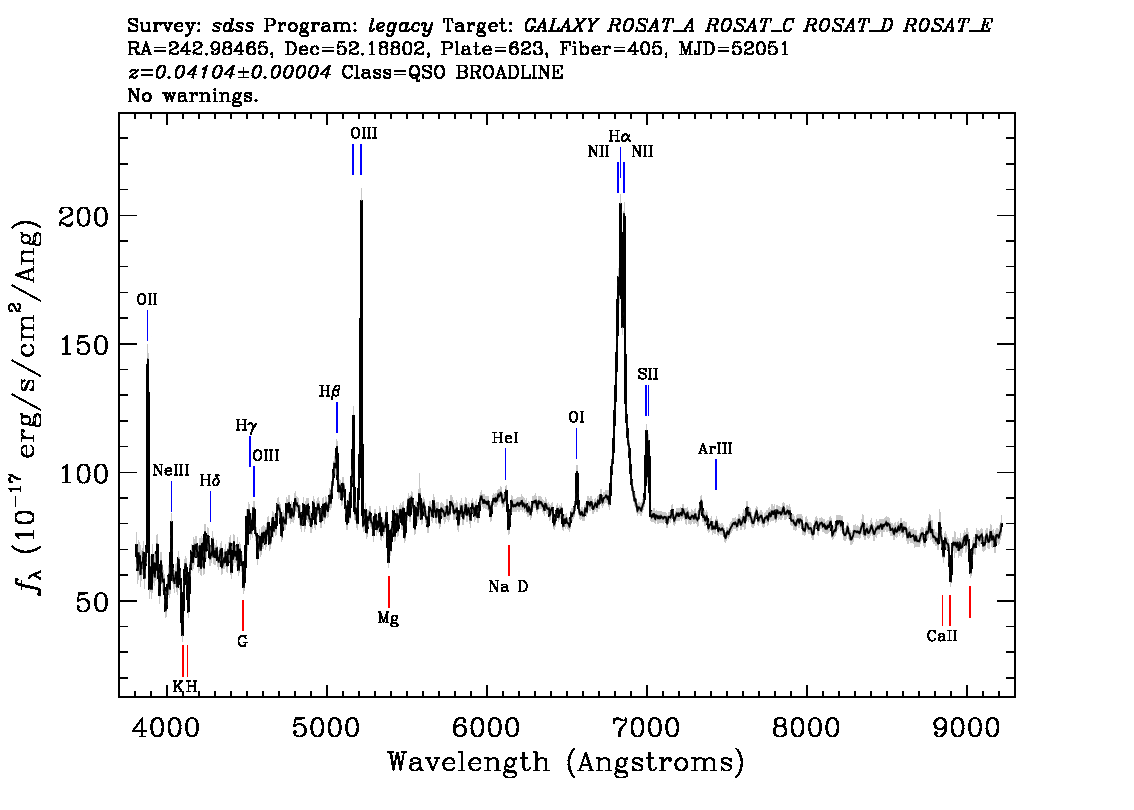}
    \end{subfigure}
    \begin{subfigure}[b]{0.32\textwidth}
        \includegraphics[width=\textwidth]{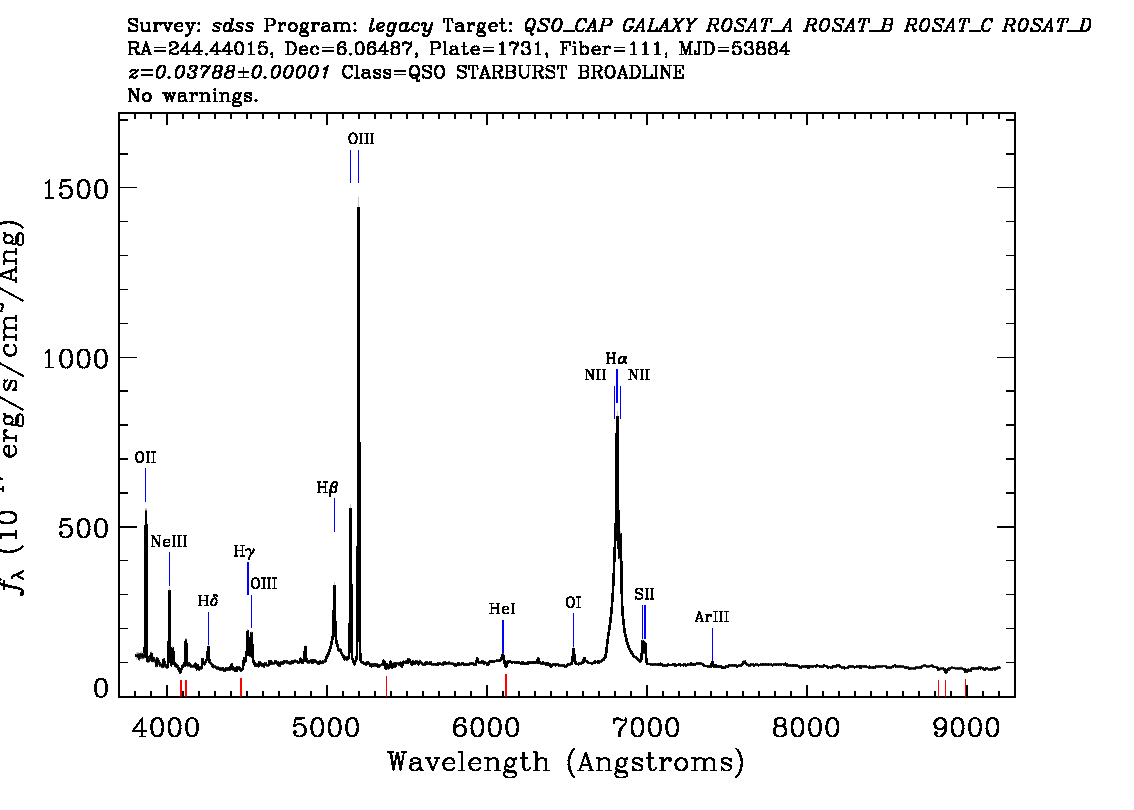}
    \end{subfigure}
    \begin{subfigure}[b]{0.32\textwidth}
        \includegraphics[width=\textwidth]{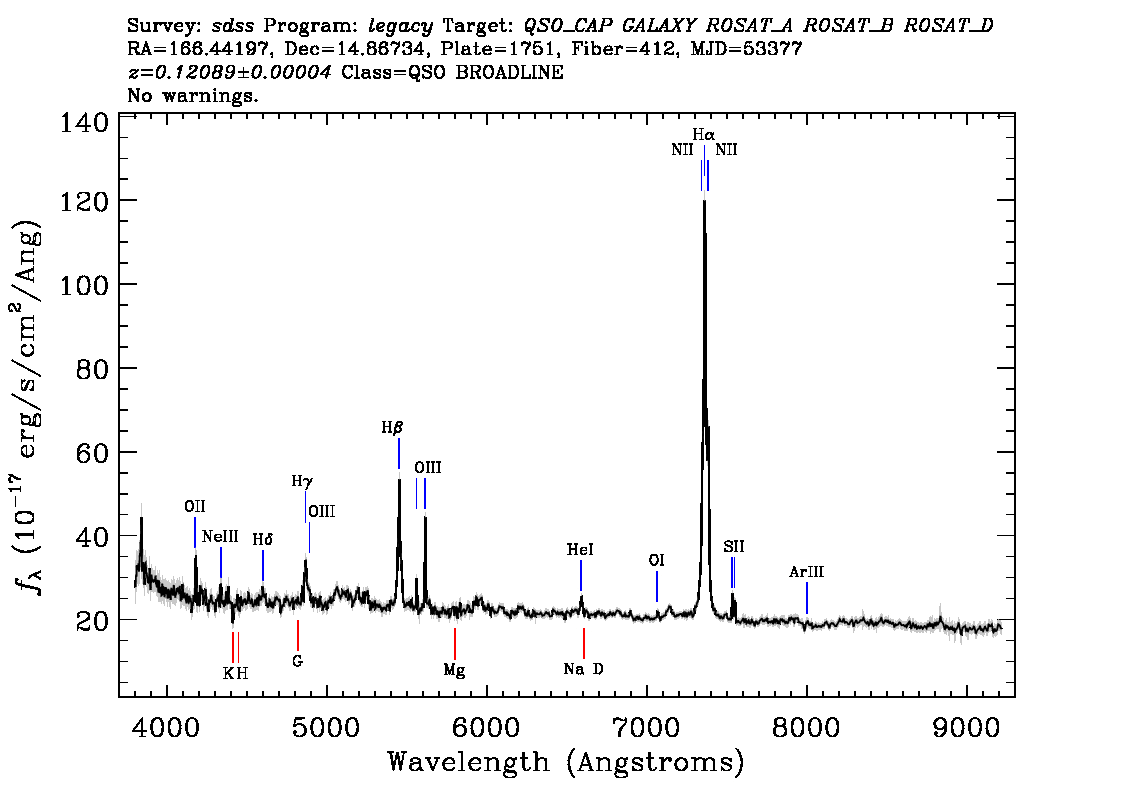}
    \end{subfigure}
    \caption{Sample of ten random optical spectra of type 1 Seyfert AGNs. The identified emission lines are indicated. These spectra were randomly picked from different
redshift ranges up to $0.2$. credits to the SDSS.}\label{fig:sy1}
\end{figure*}

\begin{figure*}
    \centering
    \begin{subfigure}[b]{0.32\textwidth}
        \includegraphics[width=\textwidth]{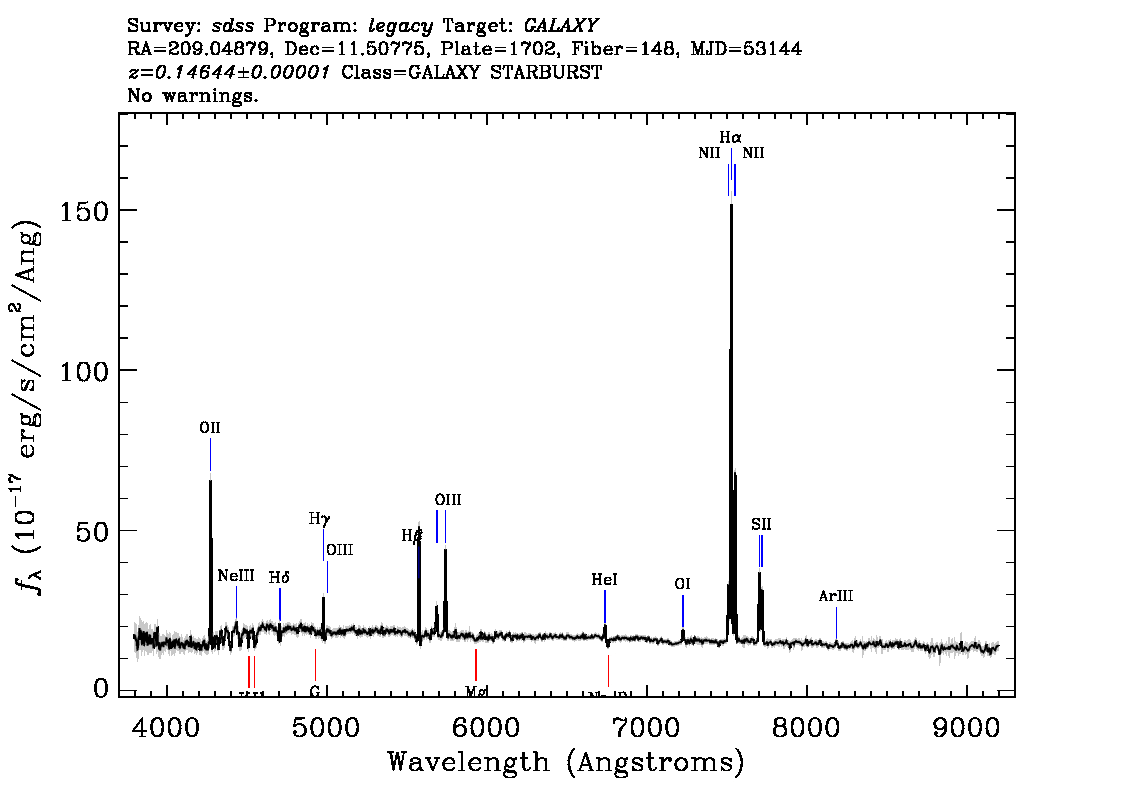}
    \end{subfigure}
    ~ 
    \begin{subfigure}[b]{0.32\textwidth}
      \includegraphics[width=\textwidth]{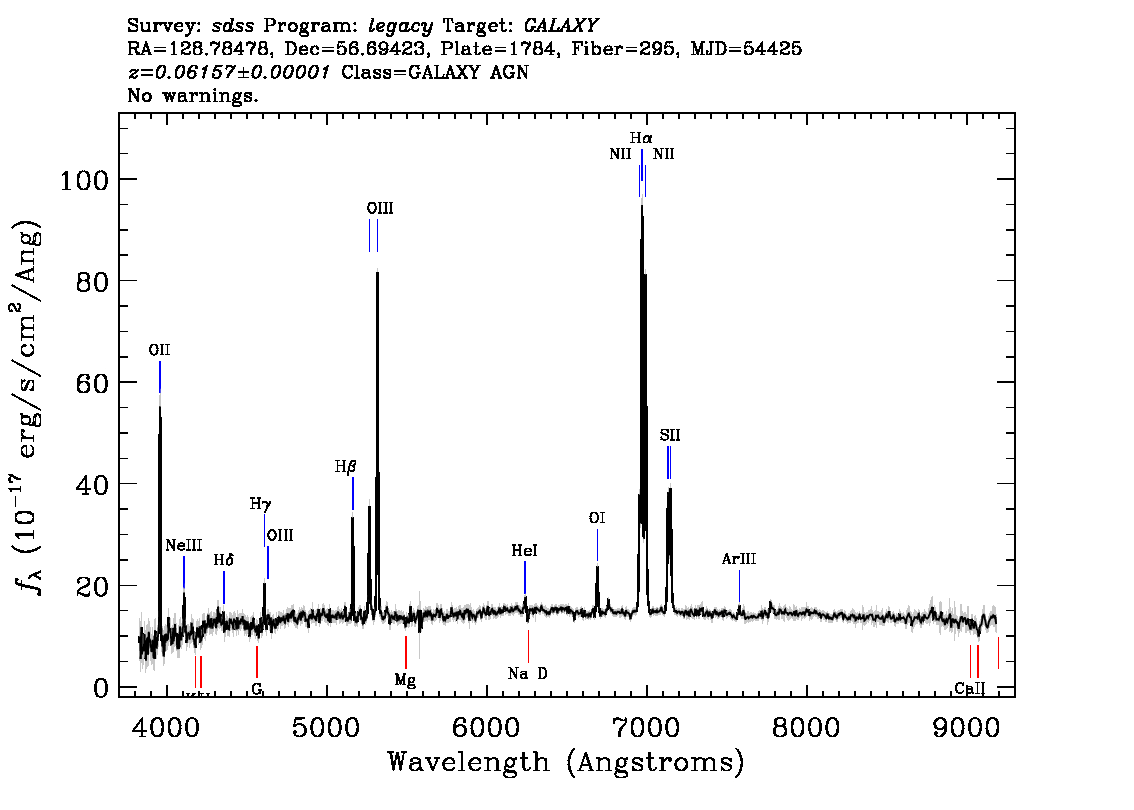}
    \end{subfigure}
    ~ 
    \begin{subfigure}[b]{0.32\textwidth}
        \includegraphics[width=\textwidth]{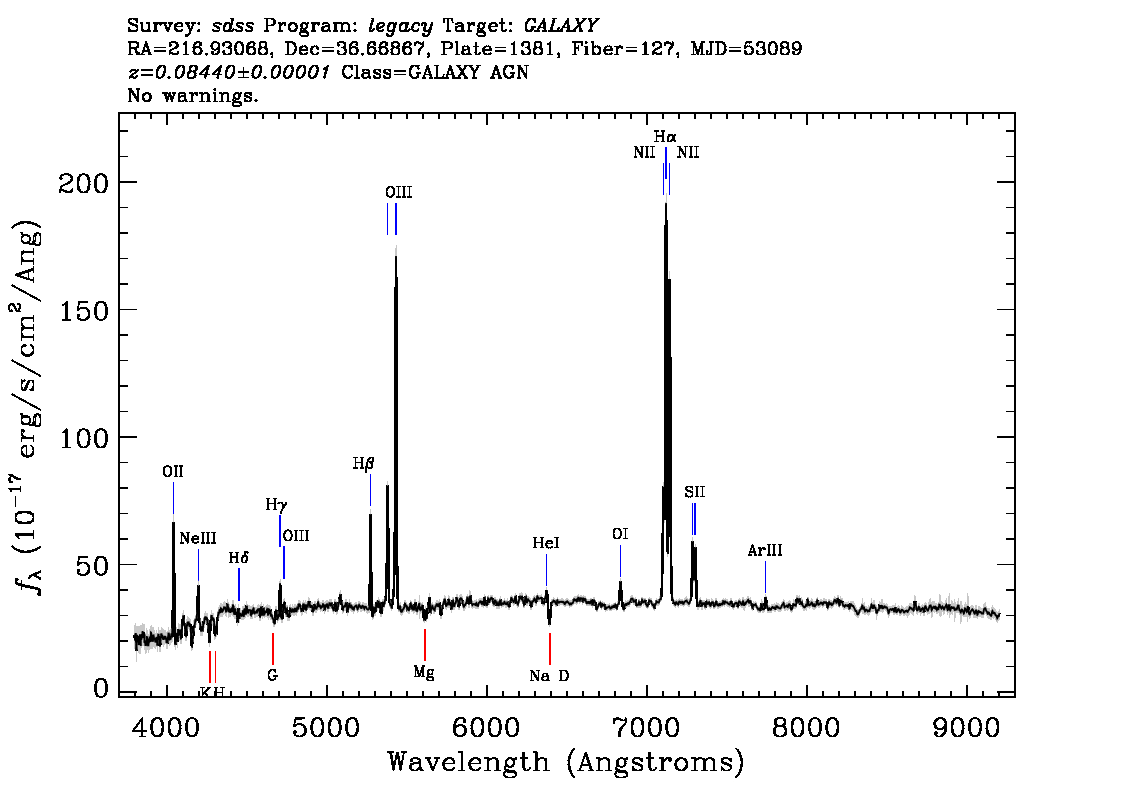}
    \end{subfigure}
    \begin{subfigure}[b]{0.32\textwidth}
        \includegraphics[width=\textwidth]{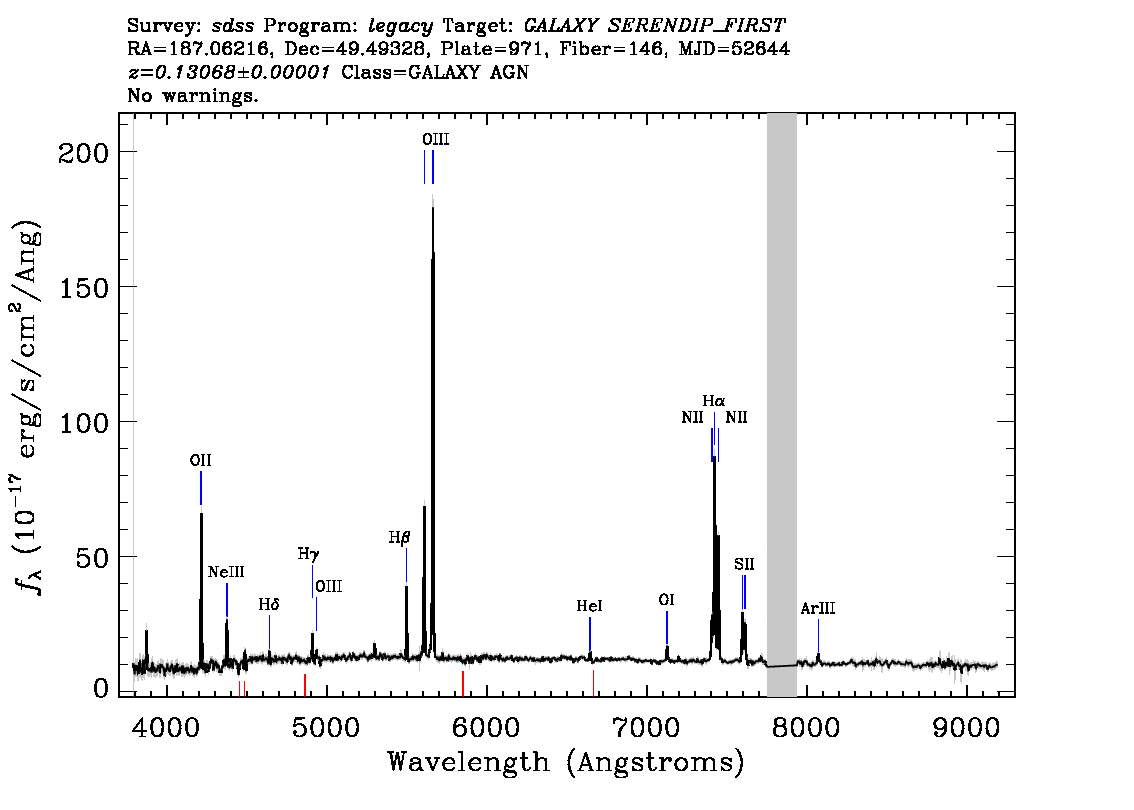}
    \end{subfigure}
    \begin{subfigure}[b]{0.32\textwidth}
        \includegraphics[width=\textwidth]{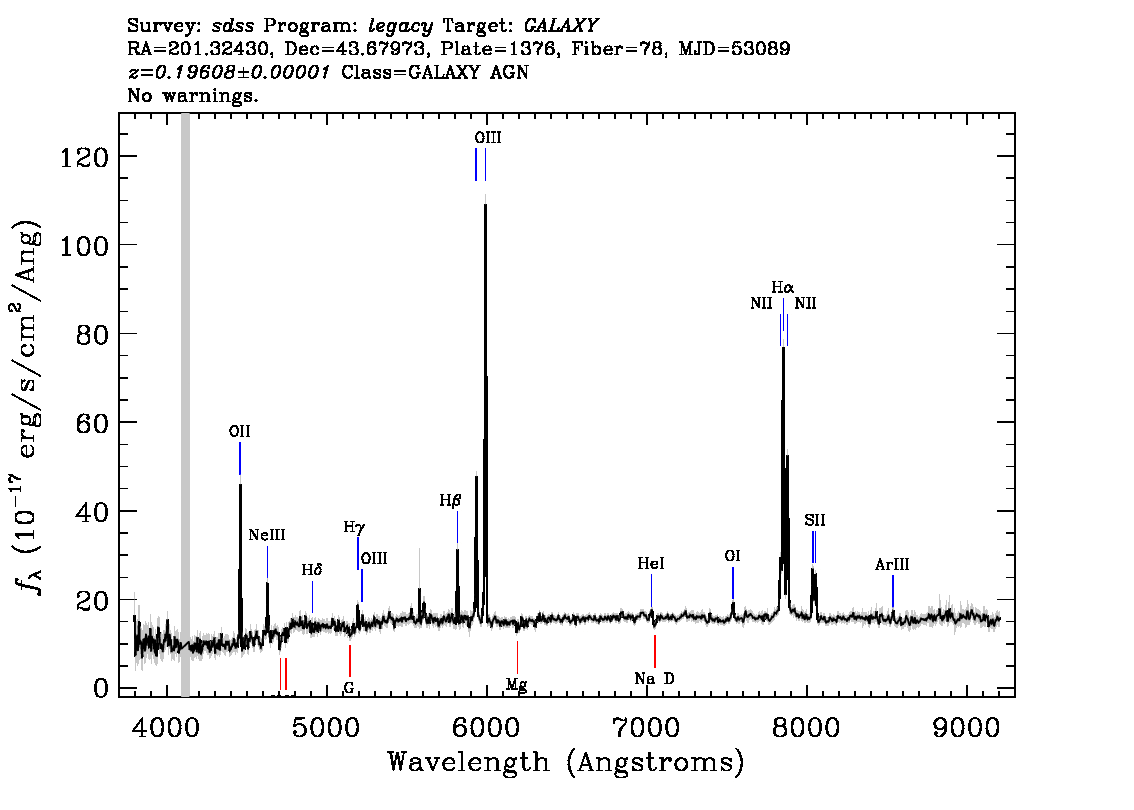}
    \end{subfigure}
    \begin{subfigure}[b]{0.32\textwidth}
        \includegraphics[width=\textwidth]{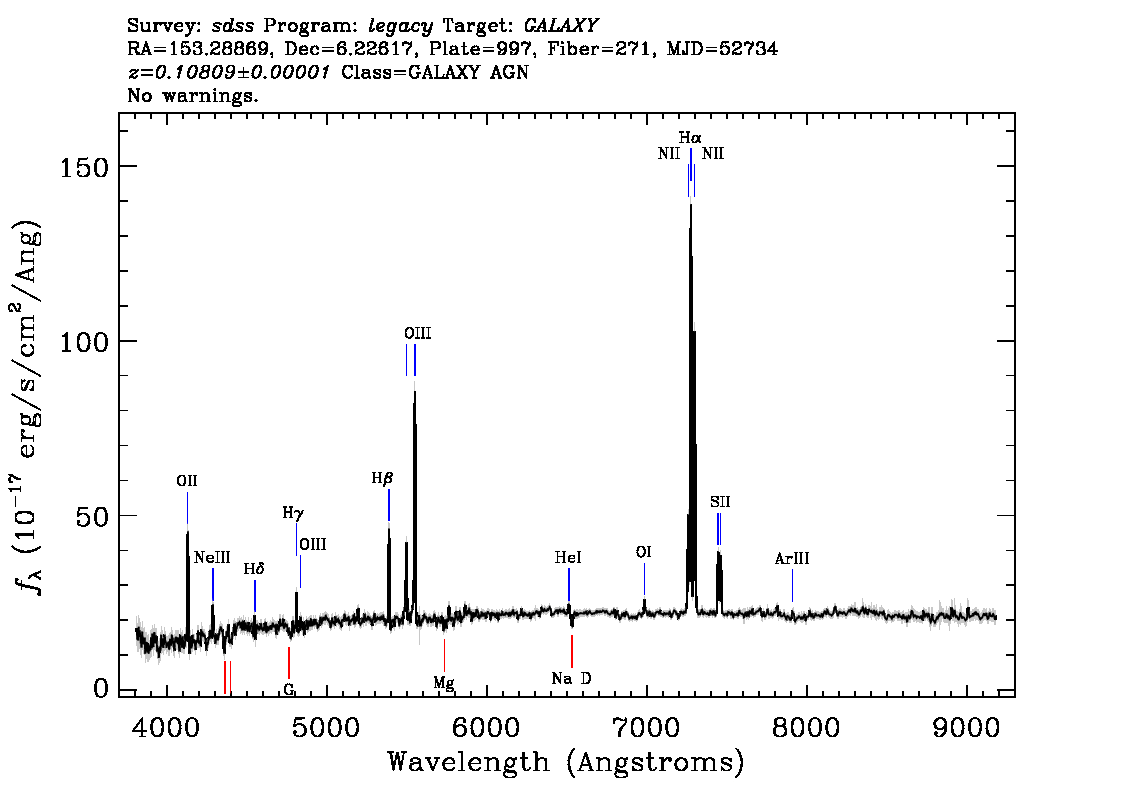}
    \end{subfigure}
    \begin{subfigure}[b]{0.32\textwidth}
        \includegraphics[width=\textwidth]{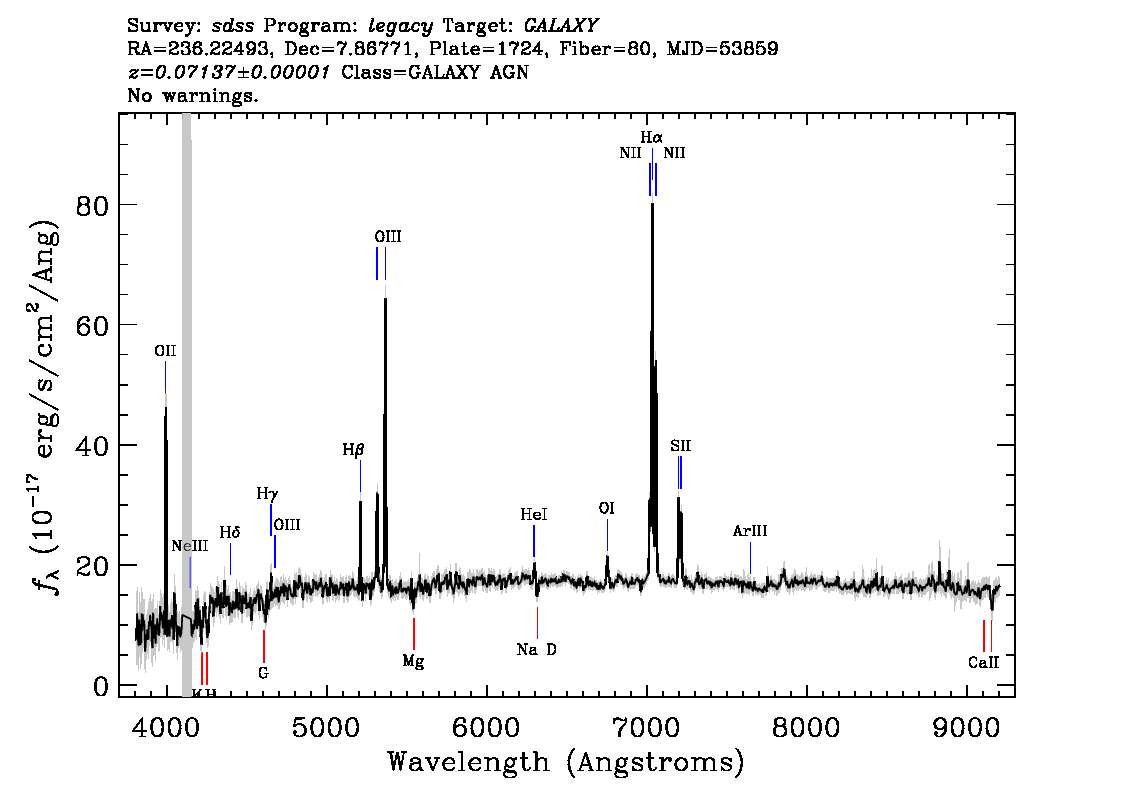}
    \end{subfigure}
    \begin{subfigure}[b]{0.32\textwidth}
        \includegraphics[width=\textwidth]{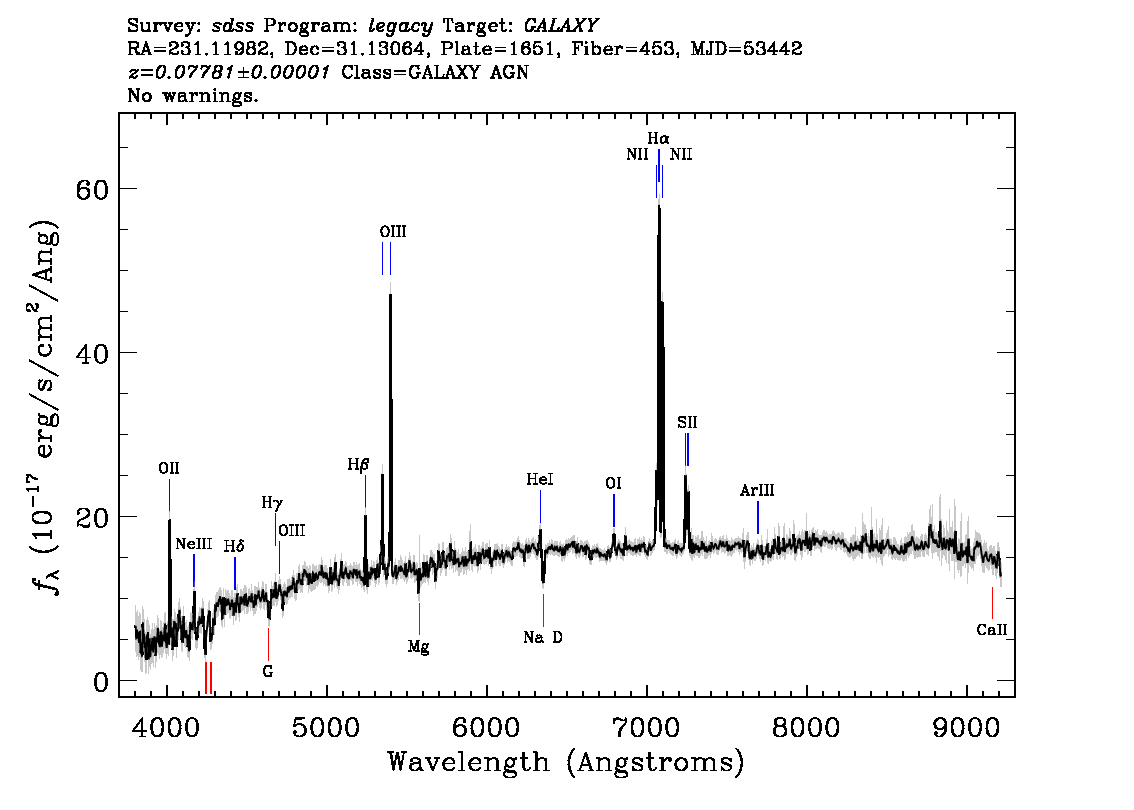}
    \end{subfigure}
    \begin{subfigure}[b]{0.32\textwidth}
        \includegraphics[width=\textwidth]{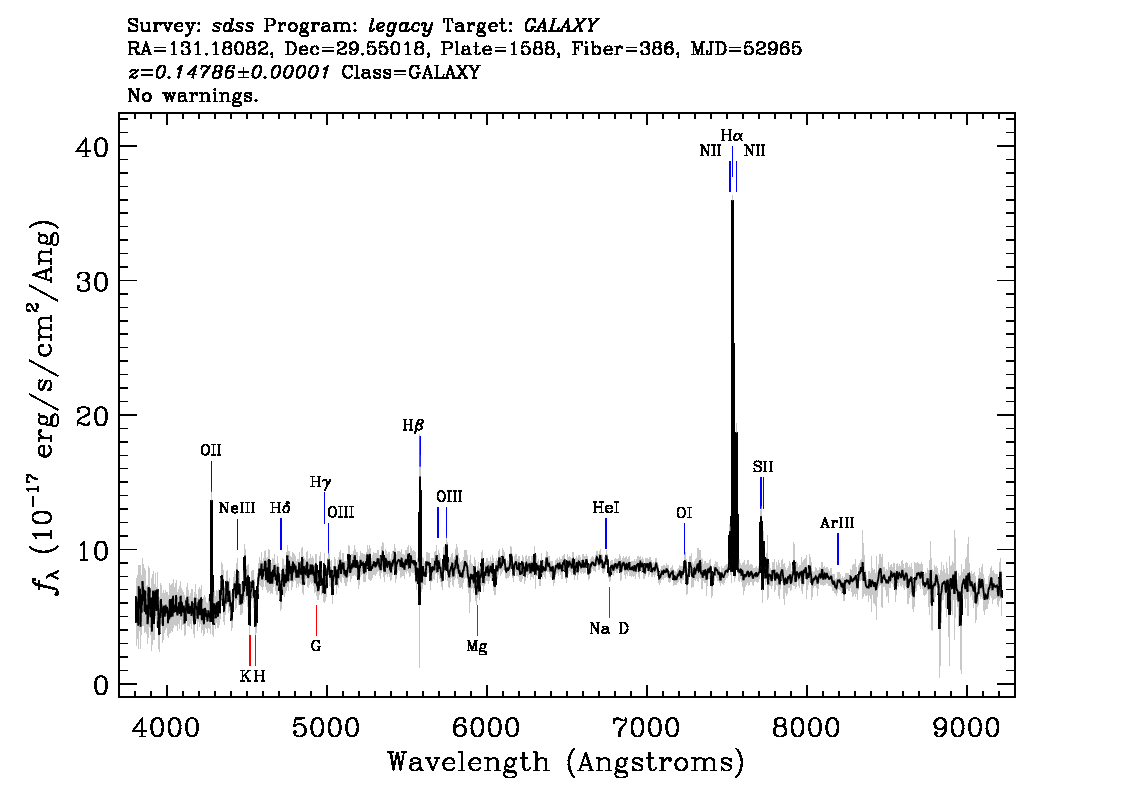}
    \end{subfigure}
    \begin{subfigure}[b]{0.32\textwidth}
        \includegraphics[width=\textwidth]{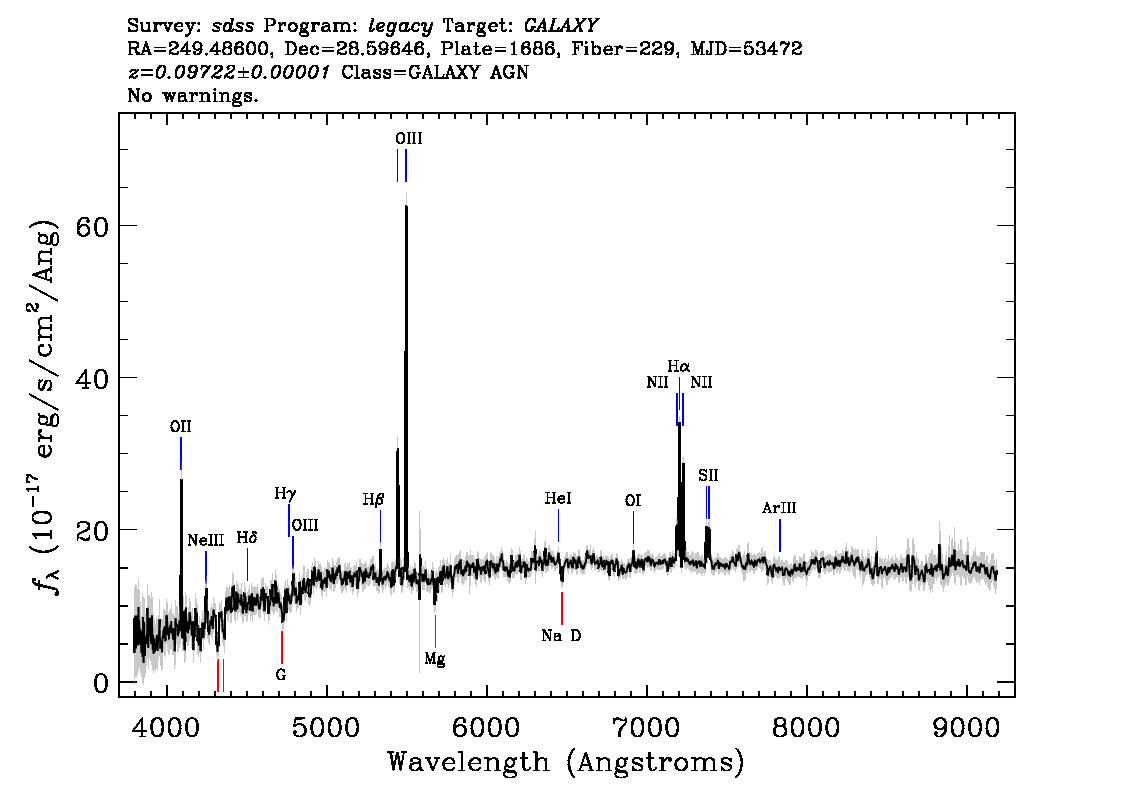}
    \end{subfigure}
    \caption{Sample of ten random optical spectra of type 2 Seyfert AGNs. The identified emission lines are indicated. These spectra were randomly picked from different
redshift ranges up to $0.2$. credits to the SDSS.}\label{fig:sy2}
\end{figure*}

The authors, in order to demonstrate that the final Sy1 sample consists of well-selected type 1 AGNs, present ten random Sy1 optical spectra (\ref{fig:sy1}), limited to $z<0.2$, from the SDSS database. On the other hand,  they comparatively present a sample of ten random Sy2 optical spectra (\ref{fig:sy2}) with redshifts up to $z=0.2$. 


\end{appendix}

\end{document}